\documentclass[preprint,12pt]{elsarticle}

\usepackage{graphicx} 
\usepackage[utf8]{inputenc}
\usepackage{url}
\usepackage{booktabs}
\usepackage[table,xcdraw]{xcolor}
\usepackage{enumerate}
\usepackage{float}
\usepackage{multirow}
\usepackage{adjustbox}
\usepackage{amsmath, amsthm, amssymb}
\usepackage{hyperref}

\usepackage{todonotes}

\newtheorem{prop}{Proposition}
\newtheorem{obser}{Observation}
\theoremstyle{definition}
\newtheorem{definition}{Definition}
\newtheorem{example}{Example}

\journal{Applied Soft Computing}
\date{July 27, 2023}

\newcommand{\czm}[1]{\textcolor{black}{#1}}
\newcommand{\czmrev}[1]{\textcolor{black}{#1}}
\newcommand{\asocrev}[1]{\textcolor{black}{#1}}
\newcommand{\nzmrev}[1]{\textcolor{black}{#1}}
\newcommand{\revdos}[1]{\textcolor{black}{#1}}

\begin{document}
\begin{frontmatter}

\title{\czmrev{Design and consensus content validity of the questionnaire for b-learning education: A 2-Tuple Fuzzy Linguistic Delphi based Decision Support Tool}}

\author[label1]{Rosana Montes}
\author[label2]{Cristina Zuheros\corref{cor1}}
\author[label1]{Jeovani Morales}
\author[label1]{Noe Zermeño}
\author[label1]{Jeronimo Duran}
\author[label2,label3]{Francisco Herrera}
\cortext[cor1]{Corresponding author.\\
E-mail addresses: rosana@ugr.es (R. Montes), czuheros@ugr.es (C. Zuheros), jeovani@correo.ugr.es (J. Morales), nzermeno@correo.ugr.es (N. Zermeño), jeronimoduran@correo.ugr.es (J. Duran), herrera@decsai.ugr.es (F. Herrera).}

\address[label1]{Department of Software Engineering, Andalusian Research Institute in Data Science and Computational Intelligence (DaSCI), University of Granada, 18071, Granada, Spain.}
\address[label2]{Department of Computer Science and Artificial Intelligence, Andalusian Research Institute in Data Science and Computational Intelligence (DaSCI), University of Granada, 18071, Granada, Spain.}
\address[label3]{Faculty of Computing and Information Technology, King Abdulaziz University, Jeddah 21589, Saudi Arabia.}

\markboth{R. Montes}{A 2-Tuple Fuzzy Delphi method}


\begin{abstract} 

Classic Delphi and Fuzzy Delphi methods are used to test content validity of data collection tools such as questionnaires. Fuzzy Delphi takes the opinion issued by judges from a linguistic perspective reducing ambiguity in opinions by using fuzzy numbers. We propose an extension named 2-Tuple Fuzzy Linguistic Delphi method to deal with scenarios in which judges show different expertise degrees by using fuzzy multigranular semantics of the linguistic terms and to obtain intermediate and final results expressed by 2-tuple linguistic values. The key idea of our proposal is to validate the full questionnaire by means of the evaluation of its parts, defining the validity of each item as a Decision Making problem. Taking the opinion of experts, we measure the degree of consensus, the degree of consistency, and the linguistic score of each item, in order to detect those items that affect, positively or negatively, the quality of the instrument. Considering the real need to evaluate a b-learning educational experience with a consensual questionnaire, we present a Decision Making model for questionnaire validation that solves it. Additionally, we contribute to this consensus reaching problem by developing an online tool under GPL v3 license. The software visualizes the collective valuations for each iteration and assists to determine which parts of the questionnaire should be modified to reach a consensual solution.
\end{abstract}



\end{frontmatter}

\section{Introduction}\label{section1}
Quality research must pay attention to the quality of every research process conducted~\nzmrev{\cite{SANTIAGODELEFOSSE2016142}}. Data collection procedures are key because the following activities rely heavily on this early step. 

Questionnaires are the most used data collection tools, along with interviews and observation. A desirable property for quality in a questionnaire is the ability to measure the variables for which it was designed, that it is, its validity. \emph{Content validity} is one of three main types of validity evidence. This validity can be tested with the classic Delphi and the Fuzzy Delphi methods, by taking the consensual opinion of a panel of experts, or judges~\nzmrev{\cite{saffie2016fuzzy,padilla2021social,dawood2021towards}}. 


A relatively very popular pedagogical methodology for blending learning (b-learning) \czm{based educational scenario} is known as Flipped Classroom~\cite{BERGMANN2012flip}. It is based on \emph{flipping} moments of learning, conceptual acquisition and application of knowledge, allowing students to learn theory outside the classroom, through resources provided by the teacher, mainly videos. Learning happens in different moments: face-to-face as in traditional education \emph{blended} with online activities. Thanks to technological advances that promote interaction between students, the traditional focus of education shifts from individual to collaborative approaches by using technology. Another pedagogical methodology that uses technology in education (mostly mobile) is Mobile Learning or \emph{m-Learning}~\cite{JALDEMARK2017editorial}. Both methodologies are very recent and have attracted by separate the interest of many researchers~\cite{AlEmran18,Karabulut18}. However it is a challenge to apply and to evaluate Flipped Classroom and m-Learning methodologies in combination in a \czm{b-}learning environment. It is even more because of the lack of standardized questionnaires that address both methodologies in combination. 

We were aware of this difficulty after finding our own need. In the context of Computer Science studies in the University of Granada, a pilot experience Flipped Classroom \& m-Learning was carried out in a first-year course. We want to evaluate the experience of virtual communication between students-students and teacher-students using the Telegram\footnote{Telegram is a messaging and VOIP platform \url{https://telegram.org}} app via a questionnaire but it should be test for content validity before its application with the students. 

There is a risk of having items or questions that do not target the dimension of interest, or they are badly wording (easy to misread), or they are simply not helpful. It is therefore desirable to test a questionnaire by individually test the items that comprise it. The content validation by expert judgments can be considered as a Multi-Expert Multi-Criteria Decision Making problem\czm{, in which a group of experts evaluate alternatives with regard to some criteria. It can be improved carrying out a consensus process \cite{tan2021cyclic}.}

The application of the Computing with Words methodology in Decision Making~\czmrev{\cite{mendel2010computing,zadeh2012computing}} has made possible during decades to incorporate linguistic concepts into applied intelligent computer systems~\cite{Montes15,Montes18,morente18}. By nature it is easier for humans to give opinion in natural language than in numerical language. \textcolor{black}{For instance, when experts try to evaluate \czmrev{through} questionnaires the \emph{usability} of a website, terms such as \emph{good}, \emph{very good} or \emph{bad} are generally used~\cite{ORFANOU2015perceived} rather than numerical assessments $usability=0.8$. To the best of our knowledge,} there is little literature involving questionnaires and Linguistic Decision Making problems, though in \cite{CARRASCO2011linguistic} a Linguistic Decision Making scenario was set to normalize the results of various questionnaires in the context of different universities, allowing the comparison of the collected data between institutions.

Our proposal is \czm{to develop a content validation model} based on obtaining the linguistic opinion of judges in an iterative process for assessing reliance and consensus among the items of the instrument. \czm{Specifically, we contribute with a Decision Support System (DSS) since it is a software that assists people to take decisions \cite{aggarwal2021multi}.} \textcolor{black}{It implements} the 2-Tuple Fuzzy Linguistic Delphi method which is an extension of the Fuzzy Delphi with linguistic information represented by the 2-tuple linguistic representation model \cite{HERRERAYMARTINEZ2000}. It is used to test, \textcolor{black}{by consensus,} the content validity of a questionnaire for an experience in b-learning. This task is assisted by a software web tool which is open and free.  

In summary, \textcolor{black}{our proposal provides}:

\begin{itemize}

\item \czm{A 2-Tuple Fuzzy Linguistic Delphi model to test content validity of a questionnaire, which is a property that must be satisfied by any data collection tool. It is suitable to handle multigranular scenarios \cite{zhao2021linguistic}.} 

\item \czm{We design a consensus process to achieve a suitable degree of agreement between the expert evaluations. This process is a dynamic mechanism in which experts change their evaluations based on the consensus and reliance indexes.} To find a consensual solution in a small number of iterations, we compute many linguistic scores for each item as the results of a Multi Expert Multi Criteria Linguistic Decision Making model.

\item \textcolor{black}{\czm{We provide a web tool based DSS as an} online tool for 2-Tuple Fuzzy Linguistic Delphi method application. The moderator can freely use our proposed model. It is also an informative tool for the expert panel to visualize the degree of consensus between them.}

\item \textcolor{black}{\czm{A} case use in which we validated the questionnaire for \czm{a b-learning based educational scenario.}} After two iterations of 2-Tuple Fuzzy Linguistic Delphi method, we get the consensual questionnaire. \textcolor{black}{This questionnaire is designed specifically to measure the concept \emph{satisfaction with} the combined use of Flipped Classroom and m-Learning methodologies in Higher Education. In education, new trends and pedagogical methodologies supported by technology require the design of adapted data collection and we have get a consensual version for its application in the course \emph{Fundamentals of Software} to know the students' opinion}.

\end{itemize} 

The paper is structured as follows. Section~\ref{section2} reflects on the objectives for questionnaires validation and the design of a questionnaire to be use in b-learning environments. Section~\ref{section3} introduces \czm{the} 2-Tuple Fuzzy Linguistic method. Section~\ref{section4} describes the software tool that supports the moderator in the task of adapting the questionnaire to reach a consensual result in few iterations. Section~\ref{section5} analyzes the results of the case study that validates a questionnaire for b-learning with an expert panel of nine judges. Finally the paper is concluded in Section~\ref{section6}.

\section{\czmrev{Background}}\label{section2new}

This section presents the basic knowledge underlying the proposed 2-Tuple Fuzzy Linguistic Delphi method to questionnaire testing and validation. \revdos{The objective is to provide a valid questionnaire, so Section~\ref{sec2new1} reviews the properties that should be addressed. Our proposal is an extension of the Delphi method, which is a technique that has been used for decades, as it is explained in the literature review in Section~\ref{sec2new2}. Steps for application of Delphi are given in Section~\ref{sec2new3}. For our proposal to be implemented, an underlying linguistic operational model is required. We explain the 2-tuple linguistic model in Section~\ref{sec2new4}. To explain the domain of the assessments, the multi-granular linguistic information is depicted in Section~\ref{sec2new5}.}

\subsection{Properties of a valid questionnaire}\label{sec2new1}
\revdos{Data collection techniques, such as questionnaires, are very common in scientific research. The design of a questionnaire requires a methodical process of design and validation. Consensus methods for questionnaire validation includes the Delphi method and extensions. First, the structure must define each item in terms of description (the text that the user reads), type (open or closed question) and answering scale (yes/no answer or Likert style). Related items might be  grouped in dimensions. Secondly, it is necessary to verify the following properties.}

\revdos{A valid questionnaire possesses the following properties:}

\begin{enumerate}
\item \revdos{\textit{Reliability}. Ensures trustworthiness and accuracy of data collected. Cronbach's alpha measures internal consistency (ideally above 0.70)~\cite{PONTEROTTO2007}. Pearson's correlation index helps eliminate items with homogeneous indexes.}

\item \revdos{\textit{Objectivity}. Measures the extent to which biases and tendencies of researchers influence the questionnaire's administration, qualification, and interpretation.}

\item \revdos{\textit{Validity}. The capacity to measure the intended variable, with three sub-dimensions:} 
\begin{enumerate}
    
    \item \revdos{\textit{Criterion validity}. The effectiveness in predicting the variable of interest through validity coefficient (correlation between test and criterion)~\cite{evers2013assessing}.}
    
    \item \revdos{\textit{Construct validity}. Tests whether dimensions contribute to the overall evaluation of the questionnaire, examined using the KMO test~\cite{KAISER1974} and Barlett sphericity test~\cite{BARLETT1950}.}
    
    \item \revdos{\textit{Content validity}. Measures comprehension of questions and dimension adjustment, validated through statistical assessment and judges' validation by expert panels~\cite{ding2002assessing, hyrkas2003validating}. Judges' validation is defined as a consensus among qualified persons who can issue evidence. }

\end{enumerate}	
\end{enumerate}

\subsection{\revdos{Literature review}}\label{sec2new2}
\revdos{The Delphi method has evolved since its introduction in the 1950s, with versions falling into three categories: the `classic' Delphi for establishing facts, the `policy' Delphi for generating ideas, and the decision Delphi for making judgments. The Electronic Delphi method~\cite{Walker23} (e-Delphi), a modern adaptation, utilizes the technological capabilities and web-based form applications for filtering consensus among experts. A web-based application can leverage a larger number of experts, making e-Delphi more efficient compared to traditional methods, but this proposal still requires a moderator to evaluate the responses.}

\revdos{It is not uncommon to find evidence of the application of this expert consultation technique, applied in 2-3 rounds, to validate questionnaires to be applied in different areas of knowledge:}
\begin{itemize}
\item \revdos{In education, a questionnaire on intercultural practices in bilingual schools, aiming to identify good practices and develop intercultural education guidelines is validated~\cite{Gomez21}. The study utilizes the Delphi method to validate content and comprehension, and internal consistency is measured through Cronbach's alpha coefficient, while factor analysis ensures construct validity. Despite acknowledging limitations, such as its applicability to other research populations, the results provide valuable insights for enhancing practices in bilingual education.} 
\item \revdos{A sustainability study~\cite{Shah19} applied the Modified Delphi method to prioritize barriers hindering the use of renewable energy sources in Pakistan. Five main barrier categories were identified, with political and regulatory barriers ranking highest. Fuzzy Analytical Hierarchical Process (FAHP) was used to calculate weights and rankings of the barriers and sub-barriers.}
\item \revdos{In medicine, the study~\cite{Miller19} utilized a modified Delphi method to develop a checklist of essential supervisory behaviors for pediatric residents leading inpatient rounds. The checklist was piloted at two hospitals to facilitate real-time feedback. However, the tool is not accessible for external users. The study~\cite{Shinners21} utilized an e-Delphi method, involving experts in health and technology, to create a questionnaire measuring perceptions of Artificial Intelligence among healthcare professionals. Consensus was achieved through three rounds of online surveying and group discussions, resulting in a reliable and validated questionnaire. In the sport field, Partnet \textit{et al.} \cite{partner2023development} follows a 3-stage online Delphi process to develop a questionnaire to monitor symptoms of rugby player shoulder dysfunctions.}  
\item \revdos{In transportation, the Delphi method combined with the analytic network process (ANP), was employed to investigate the feasibility of autonomous train operation (ATO)~\cite{DelphiANP23}. Delphi questionnaires were used to identify opportunities, problems, and determinants for ATO, while the ANP method weighted these factors.}

\end{itemize}

\subsection{The Delphi method} \label{sec2new3}

The Delphi method is an iterative process used to collect and extract expert opinions using a series of questionnaires with interspersed feedback ~\cite{Okoli2004}. Each version of the questionnaire is based on the previous iteration. Consensus processes refer to how to reach the maximum degree of agreement between experts on the set of alternative solutions, \revdos{and it is used to stop the iterative process.} \revdos{To conduct the Delphi method a set of steps must be performed systematically. In short, the workflow is described below:}

\begin{description}
    \item[\revdos{\textbf{Preliminary Phase}}] \revdos{It is performed by a person, called moderator.}
    \begin{enumerate}
        \item \revdos{Identify the problem and features.}
        \item \revdos{Establish a coordination group to prepare the pilot questionnaire.}
        \item \revdos{Select a panel of experts based on expertise, reputation, availability, and impartiality.}
    \end{enumerate}

    \item[\revdos{\textbf{Assessment Phase}}] \revdos{It is an iterative process conducted by the expert panel guided by the moderator. }
    \begin{enumerate}
        \item \revdos{Disseminate the questionnaire to judges independently.}
        \item \revdos{Sort, assess, and compare responses obtained in the first iteration.}
        \item \revdos{Modify questionnaire items based on judges' suggestions.}
        \item \revdos{Disseminate the new version to judges independently.}
        \item \revdos{Provide feedback to judges on each iteration.}
        \item \revdos{Repeat steps from short (2) to feedback (5), until positive consistency or acceptable consensus is achieved.}
    \end{enumerate}
    \item[\revdos{\textbf{Consensus Phase}}] \revdos{A satisfactory level of consensus has to be reached before a solution can be obtained.}
    \begin{enumerate}
        \item \revdos{The moderator gathers suggestions and evaluations from judges.}
        \item \revdos{Generates a new version of the questionnaire incorporating them.}
        \item \revdos{Accepts or rejects suggestions.}
        \item \revdos{The expert panel assesses the modified version in subsequent iterations.}
    \end{enumerate}
\end{description}

The classical Delphi method represents high costs of application\czmrev{~\cite{AENGENHEYSTER2017}. The main issue is that achieving a satisfactory level of consensus requires multiple iterations. Usually, to speed up this technique, each item is evaluated with a binary scale} (reject or accept) with the corresponding loss of information and knowledge inherent in the expert panel. 

The Fuzzy Delphi method~\cite{MURRAY1985}, which combines fuzzy sets and the Delphi method, can be used to interpret responses linguistically and provide more reasonable results, avoiding confusion and ensuring a common understanding between expert opinions.

\revdos{To solve some of the disadvantages of the classical Delphi method we propose a linguistic perspective~\cite{ZADEH1975}, by involving fuzzy numbers as the representation of words and avoiding numbers. As expert judgments may contain ambiguity --because of the different interpretation each person may have about the items of the questionnaire-- assessments are best reflected by using qualitative values, because words are close to human reasoning. In addition of words, we offer several linguistic scales with which the expert panel can express opinions regarding the content and structure of the questionnaire.}

\subsection{The 2-Tuple linguistic representation model}\label{sec2new4}
\czmrev{A} linguistic term set $S=\{s_0,\dots,s_g\}$ \czmrev{is composed of linguistic variables. The cardinality of S, $g+1$, is usually an odd number.}  To deal with imprecision and vagueness, a linguistic term $s \in S$ is defined by a fuzzy number represented with triangular membership function uniformly distributed. Under this assumption it is guaranteed that the 2-tuple fuzzy linguistic representation model~\cite{HERRERAYMARTINEZ2000} based on symbolic translation is precise and effective, as it is a continuous representation of a linguistic term or word, and avoids loss of information in computational processes.

\begin{figure}[h]
	\centering
	\includegraphics[height=0.20\textheight] {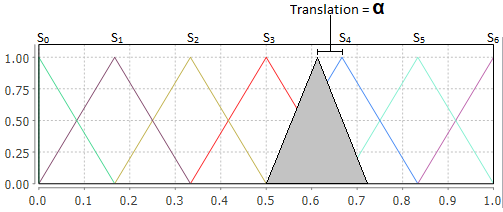}
	\caption{The \czmrev{$\alpha$ value} represents the translation of the membership function to the nearest term.}
	\label{fig:translation-alpha}
\end{figure}

\theoremstyle{definition}\label{def:tupla}
\begin{definition}~\cite{HERRERAYMARTINEZ2000} A linguistic 2-tuple $(s_i,\alpha)$ (shown in Figure~\ref{fig:translation-alpha}) is a representation of the linguistic term $s_i \in S=\{ s_0, \dots, s_g\}$ for computations in Computing with Words processes.
\begin{enumerate}
\item 
	Let $s_i \in S$ be a linguistic term whose semantic is provided by a fuzzy membership function.
\item 
	Let $\alpha \in [-0.5,0.5)$ be the value of the \emph{symbolic translation} that indicates the translation of the fuzzy membership function representing the closest term when $s_i \in S$ does not exactly match the calculated linguistic information. 
\item
	A symbolic computation operates with the indexes of the linguistic terms and obtains a value $\beta \in [0, g]$.
\end{enumerate}
\end{definition}

\begin{obser}\label{obser:zero}
The transformation of a linguistic term \czmrev{$s_i \in S$ into a 2-tuple is carried out by adding a zero as the symbolic translation to the term:}
$$ s_i \in S \longrightarrow (s_i,0) $$
\end{obser}

\begin{prop}Let $\beta \in [0,g]$ be the result of a symbolic computation. The equivalent 2-tuple is obtained my means of the function $\varDelta$ defined as:
\begin{equation}\label{eq:2Tuple}
\begin{array}{l}
\Delta: [0,g] \rightarrow S \times [-0.5,0.5)\\
\Delta(\beta)~=~(s_i,\alpha), \;with\; 
	\left\{
	\begin{array}{l}
	    s_i \;\;\;i=round(\beta), \\
	   \alpha = \beta - i
	\end{array} 
	\right .
\end{array}
\end{equation}
where \emph{round} is a function that assigns the nearest integer $i \in \{0,1,\dots,g\}$. \czmrev{There is an inverse function $\varDelta^{-1}(s_i,\alpha)~=~i + \alpha~=~\beta \in [0,g]$.}
\end{prop}

The aggregation of a set of linguistic 2-tuple values must be also a 2-tuple that summarizes this set. Many linguistic aggregation operators have been defined in the literature \cite{MARTINEZ2012, WAN2013, WEI2012,XU2011} to \czmrev{conduct} linguistic information aggregation much easier and more flexible. Let $x = \{(s_1,\alpha_1),\dots,(s_n,\alpha_n)\} = \{\beta_1,\dots,\beta_n\}$ be a set of linguistic \czmrev{2-tuple values}, \czmrev{$W=\{w_i | i=1,\dots,n\}$ a weighting vector}, and $W'$ its normalized version such as $\sum_{i=1}^n w'_i = 1$. \czmrev{The} arithmetic weighed extended mean $\bar{x}^e$ is defined as:

\begin{equation}\label{eq:avgTuples}
	\bar{x}^e (x)=\Delta\left(\frac{\sum_{i=1}^n\Delta^{-1}(s_i,\alpha_i) \cdot w_i}
	{\sum_{i=1}^n w_i} \right)
		=\Delta\left(\frac{1}{n}\sum_{i=1}^n\beta_i w'_i \right).
\end{equation}

\subsection{Multi-granular linguistic information}\label{sec2new5}

A Linguistic Hierarchy (LH)~\cite{HERRERAYMARTINEZ2001} is the union of a set of linguistic term sets, symmetrically distributed with an odd granularity of uncertainty, $n(t)$, where $t$ is a valid level of the hierarchy. 
$$
\begin{array}{l}
  LH~=~\cup_t\;\;S^{n(t)}(t)\,,\;\;t \in \{1,\dots,h\}\\
  S^{n(t)}(t)=\{s_0^{n(t)},\dots,s_{\delta_t}^{n(t)}\}\;\; where\;\;\delta_t=n(t)-1,\;\;\delta_t \in \mathbb{N}. 
\end{array}
$$

To make smooth transitions between successive levels, a term in $S^{n(t+1)}$ is the midpoint of each pair of terms belonging to the previous level $t$. Labels of this term set are known as former modal points. A set of former modal points of level $t$ is defined as:
$$
    FP_t~=~\{ fp_t^0, \dots, fp_t^{2\delta_t}\}
$$

The previous situation poses a limitation in $LH$ \czmrev{since, for example,} $S^7$ cannot be obtained from level $2$ with $S^5$ ($S^5$ can be obtained from level $1$ with $S^3$). The solution is the use of Extended Linguistic Hierarchies (ELH)~\cite{Espinilla2011}, which can manage any term set without any limitation. The ELH consist\czmrev{s} of a set of linguistic term sets $S^{n(t)}(t)$ that corresponds to a level $t$, each one with a different granularity $n(t)$. A new linguistic term set in the ELH with $t^*=m+1$ keeps all the former modal points. 
$$
   n(t^*)~=~\left(\prod_{t=1}^h\,\delta_t\right) + 1 ~=~ \delta_{t^*} + 1 
$$

It is possible to simplify $n(t^*)$ with the computation of the Least Common Multiple (LCM) value of the granularities of the family of term sets defined in the ELH. \czmrev{In this paper, to cope with general scenarios, we use $S^3$, $S^5$ and $S^7$ linguistic term sets with three, five and seven linguistic labels respectively. Thus, we get:} 
$$
	n(t^*)~=~LCM(\delta_1,\delta_2, \delta_3)+1~=~LCM(2,4,6)+1~=~13
$$

\begin{figure}[h]
\begin{center}
\includegraphics[width=8.64cm,height=8cm]{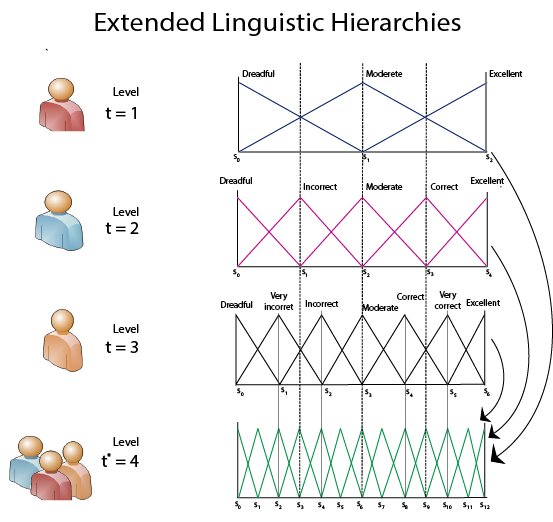}
\end{center}
\caption{\czmrev{We translate} linguistic values in $S^3$, $S^5$ or $S^7$ to level $t^*$, which in this case is $S^{13}$.}
\label{fig:ELH}
\end{figure}

The previous expression means that our computations are done under $S^{13}$, which is the bigger scale with common multiplier, as it is shown graphically at Figure~\ref{fig:ELH}. In an ELH, each formal model point $fp_{t^*}^i \in [0,1]$ is located at:
$$
  j~=~\frac{i \cdot \delta_{t^*}}{\delta_t} \to FP_t \subset FP_{t^*} \;\;\forall t = \{1,\dots,h\}
$$

\czmrev{There is a transformation function $TF_{t*}^t$ to convert} a term $s_j \in S^{n(t)}$ into the equivalent term $s_k$ expressed in $S^{n(t^*)}$ with $t < t^*$: 

\begin{equation}\label{eq:transFunct}
   TF_{t^*}^t(s_j^{n(t)}, \alpha_j)~=~\Delta\left(\frac{\Delta^{-1}(s_j^{n(t)}, \alpha_j)\;\cdot\;(n(t^*)-1)}{n(t)-1}\right)~=~(s_k^{n(t^*)}, \alpha_k)
\end{equation}

\section{The design of a questionnaire for b-learning}\label{section2}
\czmrev{This section presents a piloting experience that we have conducted in Higher Education.} We need to measure \czmrev{it} in terms of satisfaction and virtual communication. For that reason, 
 Section~\ref{section22} \czmrev{presents} a description of the methodologies applied in b-learning based educational scenarios. \czmrev{Particularly, we apply Flipped Classroom and m-Learning methodologies into the experience} in the course \emph{Fundamentals of Software}\czmrev{, as shown in Section~\ref{section23}.} Finally, Section~\ref{section24} \czmrev{presents a} questionnaire \czmrev{that allows students to evaluate the experience and which has to be validated by the proposed 2-Tuple Fuzzy Linguistic Delphi model.}

\subsection{Methodology and research question}\label{section22}
\czmrev{New technologies offer a multitude of new opportunities for teaching and learning.} Particularly in Higher Education\czmrev{,} students must be able to self-manage their learning processes and should be able to communicate effectively through the network. Current communication tools facilitate interaction and collaboration in virtual spaces. The use of \emph{flipped classroom} and \emph{m-learning} methodologies helps to overcome the distance between teachers and students and can improve the learning outcomes.

\czmrev{Flipped classroom~\cite{BERGMANN2012flip} is a methodology in which learning takes place in-classroom and out-of-classroom.} The learning activities are undertaken outside the classroom through resources provided by the teacher\czmrev{, such as activities to be solved inside and outside the classroom,} in a collaborative and meaningful way with the support of a facilitator\czmrev{, who can be the teacher or a tutor. The main objective of flipped classroom is} to promote more active and responsible learning on the part of students~\nzmrev{\cite{awidi2019impact}}.

\czmrev{Mobile Learning or m-Learning~\cite{JALDEMARK2017editorial} is a methodology that facilitates} the communication regardless of the time, devices and geographical location of the participants in the teaching-learning process\czmrev{~\cite{AZNARDIAZ2020}}. \czmrev{It} seeks to respond to the educational demand of the 21st century by providing advantages such as customizing learning experiences, achieving meaningful learning and developing professional skills \czmrev{~\cite{Irwanto2023}}.

Collaboration and virtual communication are fundamental aspects of b-learning because of the effect they have on learning and satisfaction~\cite{KIM2011763}. The use of a communication tool and an scenario of collaboration and communication between students, and between the teacher and the students, is the core of our piloting experience. The question that this research is intended to answer is: \emph{How can we promote virtual communication and satisfaction in Higher Education when flipped classroom and m-Learning methodologies are applied?}

The theoretical model underlying the learning community is the \emph{Community of Inquire} (CoI) model ~\cite{GARRISON201384}. \czmrev{There are researches} regarding data collection instruments for the CoI framework~\cite{Arbaugh08} \czmrev{but} flipped classroom and m-Learning methodologies are not implicit in the design. According to the \czmrev{CoI model}, in the communication that takes place in a virtual community there are three styles of presence or core elements: 
\begin{itemize}
	\item \textit{Cognitive Presence}: \czmrev{It allows students to construct new educational experiences through series of phases.}
	\item \textit{Social Presence:} \czmrev{It develops} interpersonal relationships through the media available in the learning environment.
	\item \textit{Teaching Presence:} \czmrev{It integrates} the above elements through design, direct teaching and resource facilitation. \czmrev{Anyone can play this role, it does not refer exclusively to the teacher or tutor.}
\end{itemize}

\subsection{Experience description}\label{section23}
A b-learning experience has been conducted \czmrev{in a Higher Education context since the academic year 2017/2018 until present day.} The selected subject is \emph{Fundamentals of Software}\czmrev{, taught by Rosana Montes to four groups in the} first year of the Degree in Computer Engineering of the University of Granada. The experience puts in practice b-learning elements by applying combined methodologies of Flipped Classroom and m-Learning with the support of technologies such as Moodle and Telegram. \czmrev{We focus on out-of classroom sessions because it represents the most innovative part of our proposal and in-classroom sessions are conducted by traditional teaching.}

\czmrev{The \emph{Moodle} platform\footnote{Moodle \url{https://moodle.org}} was used to share course resources, such as documents, glossaries, quizzes, videos, other activities, and grades, with the students. The teacher created and uploaded eleven videos presenting basic concepts of the subject in order to strengthen out-of-classroom sessions based on the flipped classroom methodology. Individual and team activities were created to put into practice the concepts learned in the videos.} Feedback to students and solutions to activities are given using this platform \czmrev{and the one we present below}.

\begin{figure}
	\centering
	\includegraphics[height=0.60\textheight] {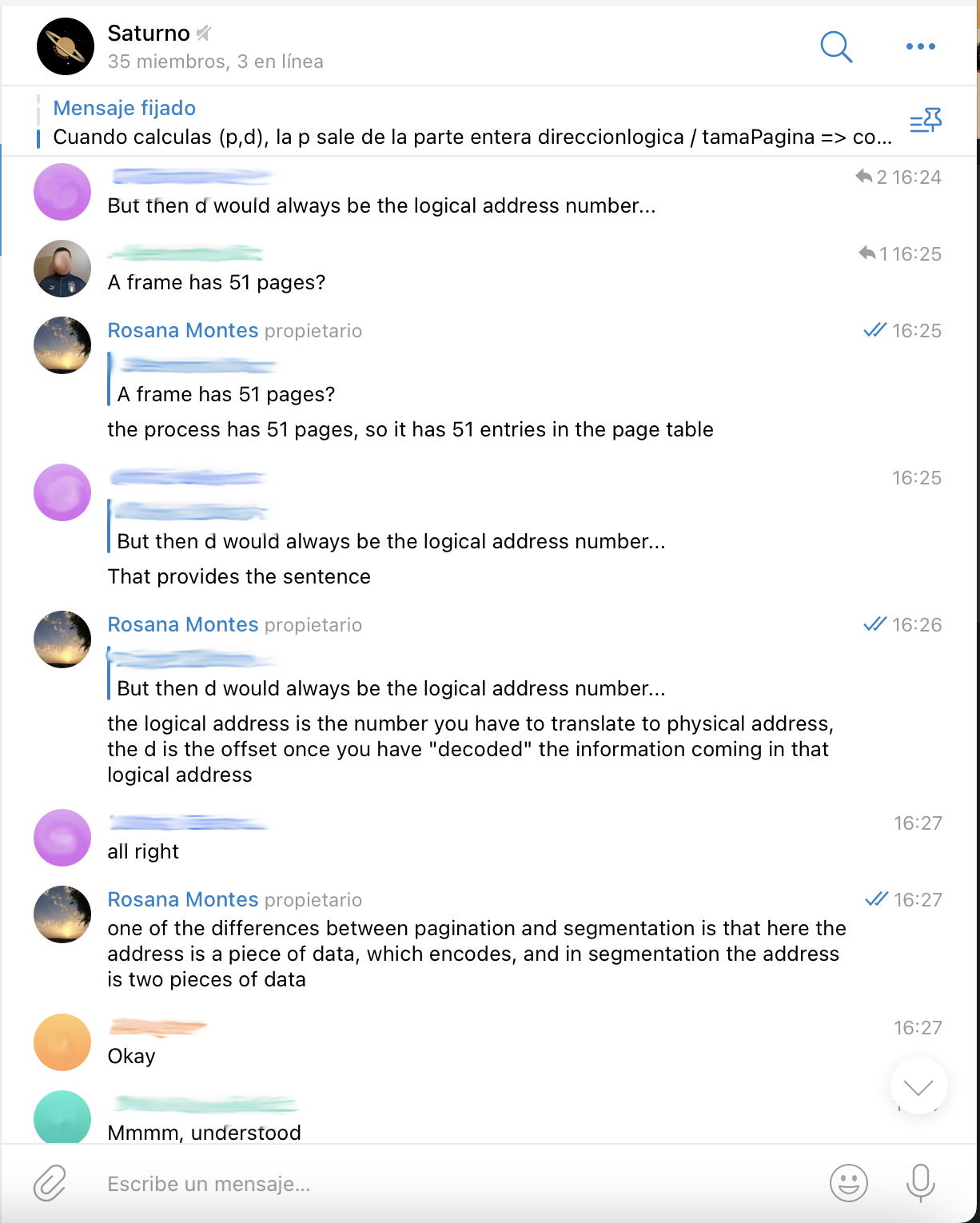}
	\caption{\textcolor{black}{Students actively participate with the teacher during the meetings.}}
	\label{fig:meeting-urano}
\end{figure}

\czmrev{The \emph{Telegram} app\footnote{Telegram \url{https://telegram.org}} was used as a tool for virtual communication, both synchronous and asynchronous. We accommodated approximately 80 students from each course of the subject into small messaging groups, that we called \emph{planets}, to achieve more fluid communication. Particularly, the} teacher set eight or more \emph{planets} (The Earth, Mars, Venus,...) and let \czmrev{the students distribute them freely, suggesting 12 to 20 participants per planet. Each planet is set a day and an hour to hold structured follow-up sessions as microblogging, that we called \emph{meetings}. Figure~\ref{fig:meeting-urano} displays a short clip of communication in a meeting.} A total of seven \emph{meetings} were scheduled for each of the \emph{planets} \czmrev{with an average duration of 30 minutes approximately.} Thousands of messages were produced between September and November of each academic year since 2017.

\czmrev{The communication scheme we employ involves both student-to-student and teacher-to-student communication and} considers unidirectional and bidirectional channels, as shown in Table~\ref{tbl:comType}. \czmrev{Asynchronous} activities follow a traditional scheme of delivering materials through a learning management system such as Moodle\czmrev{, while synchronous activities are carried out as bidirectional communication channels using Telegram. The teacher-to-student communication via Telegram is limited to the meetings. However, in some cases, the students needed to catch the attention of the teacher by mentioning her username} and she gave punctual answers. Most of the time, students were autonomous and free to communicate, provided they maintained a code of honor and good conduct. \czmrev{We require a} questionnaire that measures the virtual communication and the satisfaction with this specific experience\czmrev{, so the next section presents a questionnaire definition.}

\begin{table}[h]
	\begin{center}
		\begin{adjustbox}{max width=\textwidth}
		\begin{tabular}{c|c|c|c}
			\hline \hline 
			Type of communication & Communication flow & Activity & Tool \\ 
			\hline 
			\hline 
			\multirow{3}{*}{Asynchronous}
			& Students $\Longrightarrow$  Teacher  & Sign up for a planet & Moodle \\			
			& Teacher $\Longrightarrow$  Students & Self-managed work with videos & Moodle \\ 
			& Students $\Longleftrightarrow$  Students & Informal team-managed work & Telegram\\
			\hline 
			\multirow{2}{*}{Synchronous}
			& Students $\Longleftrightarrow$ Students & Informal team-managed work & Telegram \\ 
			& Teacher $\Longleftrightarrow$ Students & Formal follow-up meetings & Telegram \\ 
			\hline \hline 
		\end{tabular}
	\end{adjustbox}
	\caption{Synchronous and asynchronous virtual communication experiences are designed.}
	\label{tbl:comType}
	\end{center}
\end{table}

\subsection{Questionnaire definition}\label{section24}
\czmrev{The flipped classroom and m-learning experience described in Section ~\ref{section23} is valued by the students through a satisfaction questionnaire}, with which we also want \czmrev{to detect the three elements of the CoI communication model.} It is not our aim to carry out an analysis of the content of the messages, but rather to measure the degree of appearance of each presence of the CoI model~\cite{Arbaugh08}, along with the satisfaction with each presence and the general experience. \czmrev{It is easy to find questionnaires to evaluate flipped classroom and m-learning experiences separately}, but it is most harder to find questionnaires for a combined use of both methodologies.

In the literature, we only found a questionnaire specifically designed for a combined use of both methodologies~\cite{GARCIA201715}. This questionnaire tries to measure the satisfaction and communication in the underline CoI model, \czmrev{so it is adequate. Table~\ref{tbl:dimBL} shows the main elements of such questionnaire, that we note by $Q_0$.} The subscript 0 indicates that it is its initial version. It has $l=7$ dimensions and $n=45$ items. \czmrev{The present b-learning experience} poses an additional challenge since experts in flipped classroom may not be experts in m-learning field. Particularly, we even needed experts in the CoI model. Thus, a validation method based on experts judgments should consider to use a very diverse expert panel and reflect the degree of expertise of each expert and for each dimension. 
 \begin{table}[h]
	\centering
	\resizebox{12.5 cm}{!} {
		\begin{tabular}{|
				>{\columncolor[HTML]{C0C0C0}}c |c|c|c|c|c|c|c|}
			\hline
			\multicolumn{8}{|c|}{\cellcolor[HTML]{C0C0C0}Questionnaire to evaluate a piloting experience in the Degree in Computer Engineering}                                                                                      \\ \hline
			{\color[HTML]{000000} Blocks}     & \multicolumn{3}{c|} {\begin{tabular}[c]{@{}c@{}}Virtual \\ Communication 
			\end{tabular}}       & \multicolumn{4}{c|} {\begin{tabular}[c]{@{}c@{}}Students' \\ Satisfaction	
			\end{tabular}}                       \\ \hline
			{\color[HTML]{000000} Dimensions} 
			& {\begin{tabular}[c]{@{}c@{}}Cognitive \\ Presence\end{tabular}} 
			& {\begin{tabular}[c]{@{}c@{}}Social \\ Presence\end{tabular}} 
			& {\begin{tabular}[c]{@{}c@{}}Teaching \\ Presence\end{tabular}} 
			& {\begin{tabular}[c]{@{}c@{}}Cognitive \\ Presence\end{tabular}} 
			& {\begin{tabular}[c]{@{}c@{}}Social \\ Presence\end{tabular}} 
			& {\begin{tabular}[c]{@{}c@{}}Teaching \\ Presence\end{tabular}} 
			& {\begin{tabular}[c]{@{}c@{}}General \\ Satisfaction\end{tabular}} \\ \hline
				{\color[HTML]{000000} Items} 
			& {\begin{tabular}[c]{@{}c@{}}$I_1$ - $I_8$ \end{tabular}} 
			& {\begin{tabular}[c]{@{}c@{}}$I_9$ - $I_{14}$  \end{tabular}} 
			& {\begin{tabular}[c]{@{}c@{}}$I_{15}$ - $I_{21}$  \end{tabular}} 
			& {\begin{tabular}[c]{@{}c@{}}$I_{22}$ - $I_{28}$  \end{tabular}} 
			& {\begin{tabular}[c]{@{}c@{}}$I_{29}$ - $I_{35}$  \end{tabular}} 
			& {\begin{tabular}[c]{@{}c@{}}$I_{36}$ - $I_{41}$  \end{tabular}} 
			& {\begin{tabular}[c]{@{}c@{}}$I_{42}$ - $I_{45}$  \end{tabular}} \\ \hline
			
		\end{tabular}
	}
	\caption{Blocks, Dimensions and Items corresponding to the questionnaire to evaluate Virtual Communication and Students' Satisfaction in FC and m-learning methodologies.}
	\label{tbl:dimBL}
\end{table}

 \czmrev{We intend to model the validation of the questionnaire as a decision making problem, thus we present it in an standard way using mathematical notation. This way, any other valid questionnaire could be considered for evaluation through our proposed 2-Tuple Fuzzy Linguistic Delphi method. Suppose} that a given investigation covers $l$ constructs or dimensions and that a questionnaire with $n$ items has been designed to evaluate those constructs. \czmrev{Let us also assume that it is} a closed questionnaire \czmrev{so} each item is a text composed of two parts: the wording of the question and the scale to be used for the answer. Thus, a questionnaire $Q$ is a succession of items $I = \{I_1,\dots,I_r,\dots,I_n\},$ $r \in \{1,\dots,n\}$ grouped by $l$ dimensions $D_i$, $i \in \{1,\dots,l\}$:
$$
	Q~=~\{D_1,\dots,D_l\}~=~\{[I_1,I_i],[I_{i+1},I_j],[I_{j+1},I_u],\dots,[I_v,I_n]\}
$$

\section{\czmrev{The} 2-Tuple Fuzzy Linguistic Delphi method}\label{section3}
\czmrev{This section defines the proposed} 2-Tuple Fuzzy Linguistic Delphi method to be used to test validation of construct of a given questionnaire. \czmrev{Section ~\ref{section32} introduces its} underlying linguistic representation. Section~\ref{section33} \czmrev{presents the 2-Tuple Fuzzy Linguistic Delphi method workflow for evaluating questionnaires}. Section~\ref{section34} \czmrev{explains hows to solve the Multi-Expert Multi-Criteria Decision Making problems that underlying the evaluation of each item of the questionnaire.} Finally, Section~\ref{section35} \czmrev{}{depicts} the consensus model that tries to find the agreement between judges.

\subsection{The linguistic representation model}\label{section32}

\czmrev{We} implement a Computing with Words based linguistic Multi-Expert Multi-Criteria Linguistic Decision Making system, \czmrev{so} a model for linguistic data representation have to be chosen. It is our interest to provide a flexible \emph{2-tuple fuzzy linguistic delphi} method to be used by the \czm{field experts in social science}: (1) as a tool for \czm{an expert} to validate a questionnaire and, (2) as an informative tool for the \czm{expert} that, as an expert, has to reach an appropriate degree of consensus with others. We foreseen to incorporate the following characteristics:

\begin{itemize}
\item
	The iterative nature of the Delphi technique force to understand the results of the previous iteration. The collective opinion computed by a Multi-Expert Multi-Criteria Linguistic Decision Making model should be a word which is easier to understand that statistical measures such as the standard deviation or the KMO values, because words are close to human way of reason. Linguistic outputs are obtained thought the use of the 2-tuple fuzzy linguistic model~\cite{HERRERAYMARTINEZ2000}. The better understanding of the collective opinion will favor consensus-reaching processes.
\item
	The expert can choose between several linguistic term sets the one that better suits his/her degree of expertise. Most of the times a questionnaire covers many different constructs and some constructs could be distant to the expert. For instance, when the research applies different methodologies by combination. In these situations a particular expert can have high confidence in some constructs and less in others. Nonetheless the expert evaluates the questionnaire entirety and not in some parts. We incorporate the idea of expert weights per dimension (noted as $W_{D_m}$ in Section~\ref{section34}). We also assume that if you have high knowledge in a particular field, it is better to have a richer set of terms. In this way, we allow the expert to modify his/her scale at any time. 
\end{itemize} 

\czmrev{The 2-Tuple Fuzzy Linguistic Delphi method integrates the use of the 2-tuple linguistic representation model (see Section \ref{sec2new4}) and the multi-granular linguistic information (see Section \ref{sec2new5}) to address the previous features. We} consider $S^3$, $S^5$ and $S^7$ linguistic term sets \czmrev{to} cope with general scenarios. The use of multiple linguistic scales adds an \textit{Unification step} in our decision solving scheme \czmrev{by means of Equation \ref{eq:transFunct}.} In this way computations are always conducted at level $t^*$, \czmrev{which in this case is $S^{13}$, to keep} all the formal model points and to represent any value of any linguistic term set.

\subsection{The 2-Tuple Fuzzy Linguistic Delphi method \czmrev{workflow}}\label{section33}

We propose to extend the Fuzzy Delphi method by addressing experts judgments with linguistic information, represented with Extended Linguistic Hierarchies and the 2-tuple linguistic computational model. The application of the 2-Tuple Fuzzy Linguistic Delphi method follow the steps \czmrev{from the Delphi method (see Section~\ref{sec2new3}). They} are extended according to the notation used in the definition of a questionnaire (see Section~\ref{section24}) \czmrev{as follows}:

\begin{itemize}
	\item \textbf{Preliminary phase}:
	\begin{itemize}
		\item The field experts define formally the problem to be evaluated and design the items of a questionnaire $Q_0$.
		\item The field experts select and invite experts in the area to the panel of experts $J$. Optionally the research team assign to each expert a weight value with respect to $D$, the dimensions of \czmrev{questionnaire $Q_0$}. 
		\item The research team select\czmrev{s} a member to act as a moderator.
		\item The field experts select a family of $h$ linguistic term sets, with their semantics. We propose to use $h=3$ with $S^3$, $S^5$ and $S^7$.
	\end{itemize}
	\item \textbf{Assessment phase}:
	\begin{itemize}
		\item \czmrev{The moderator} starts the 2-Tuple Fuzzy Linguistic Delphi method by distributing $Q_0$ to each expert. 
		\item The expert choose\czmrev{s} a scale $S^{n(t)}$ to assess the questionnaire on the initial iteration $Q_0$, the next iteration $Q_{1}$, and so on ($Q_\iota$ with $\iota \leq max\_iterations$). 
		\item \czmrev{The moderator} uses a Decision Support System (DSS) tool (detailed in Section~\ref{section4}) to assist the validation by consensus (described in Section~\ref{section35}).\czmrev{This figure can manipulate the parameter $epsilon$, representing the \emph{satisfactory reliance level}, at any iteration by means of the DSS tool}
		\item \czmrev{The moderator} sets a new version of the questionnaire $Q_{\iota+1}$ that incorporates \textcolor{black}{the experts' open suggestions regarding $Q_\iota$}. Then, \czmrev{we repeat} the assessment phase with this new questionnaire.
		\item The procedure stop\czmrev{s} at a maximum number of iterations or when a satisfying level of consensus is achieved. 	
	\end{itemize}
	\item \textbf{Exploitation phase}:
	\begin{itemize}
		\item \czmrev{Once the} core processes \czmrev{is finished}, the research team \czmrev{has} a complete overview of the evaluation of the questionnaire \czmrev{by visualizing the DSS online tool.}
		\item \czmrev{The last} version of the questionnaire might be used in a piloting experience to conduct statistical analyses such as: Cronbach's alpha, KMO index or Berlett's sphericity to corroborate if \emph{reliability}, \emph{validity} or \emph{objectivity} are met.
		\item \czmrev{The} questionnaire can be applied in a real study \czmrev{when} there is enough statistical confidence.
	\end{itemize}
\end{itemize}	

\begin{figure}[h!]
\begin{center}
\includegraphics[width=13.66cm]{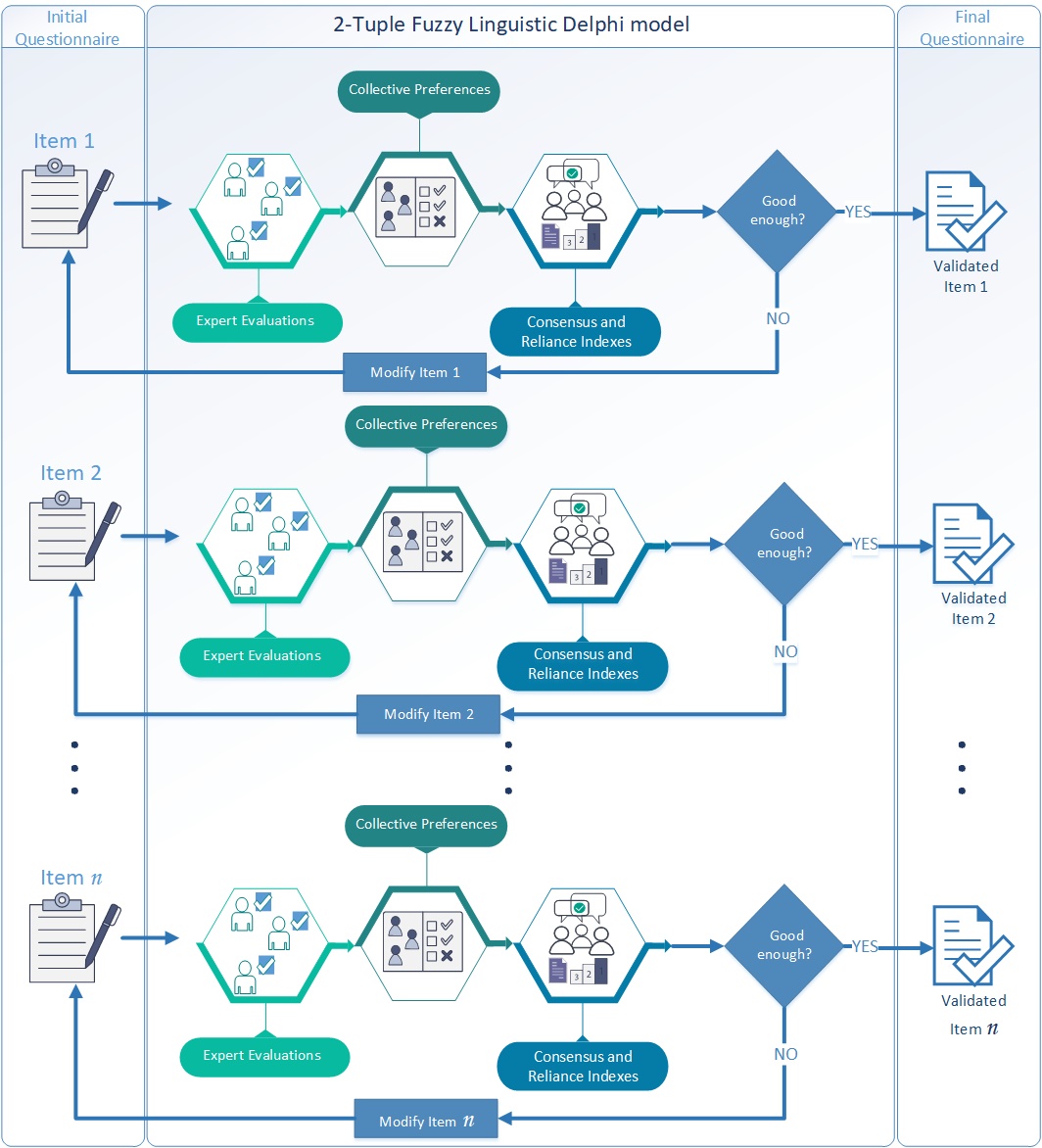}
\end{center}
\caption{\textcolor{black}{The proposed 2-Tuple Fuzzy Linguistic Delphi method solves several Multi-Expert Multi-Criteria Linguistic Decision Making problems that are repeated \czmrev{through} iterations till a consensus level is reached for each item.}}
\label{fig:general}
\end{figure}

\czmrev{Figure~\ref{fig:general} depicts the general Computing with Words scheme of the proposed 2-Tuple Fuzzy Linguistic Delphi method. It} is an iterative \czmrev{process} guided by a moderator figure \czmrev{as in} the classic Delphi method. In our approach an \czmrev{item $I_r$} of a questionnaire is accepted by consensus or rejected as a result of a Multi-Expert Multi-Criteria Linguistic Decision Making problem. \czmrev{The subsequently section presents how the model solves these problems.}

\subsection{\asocrev{The Multi-Expert Multi-Criteria Linguistic Decision Making problem}}
\label{section34}

A questionnaire $Q_\iota$ ($\iota=0, \dots, max\_iterations$) is a succession of items $I = \{I_1,\dots,I_r,\dots,I_n\}\;(r = 1,\dots,n)$ grouped by $l$ dimensions, $D$, thus: 
$$
	Q_\iota~=~\{D_1,\dots,D_l\}~=~\{[I_1,I_i],[I_{i+1},I_j],[I_{j+1},I_u],\dots,[I_v,I_n]\}
$$
We consider that to test a questionnaire of $n$ \czmrev{items}, we have to solve $n$ instances of the same Multi-Expert Multi-Criteria Linguistic Decision Making problem. This problem is defined considering the following:

\begin{itemize}
\item
	A single alternative is evaluated: the item $I_r$, \czmrev{which is composed of} its wording and its answering scale.
	
\item
	Let $J=\{J_1,\dots,J_p\} (i=1,\dots,p)$ be the expert panel involved in the validation of a questionnaire $Q_\iota$.
	
\item 
	\czmrev{The opinions are provided in} the form of a single label $s^{n(t)} \in S^{n(t)}$ with $S^{n(t)} \subset$ \czmrev{$ELH = S^3 \cup S^5 \cup S^7$}. 
	
\item
	\czmrev{The judges} can have different degrees of expertise. \czmrev{Thus,} they are rated according the expertise over each dimension by the research team. Let suppose that $I_r \in D_m$ with $m = \{1,\dots,l\}$. Then $W_{D_m} =~\{w_{1D_m},\dots, w_{pD_m}\}$ is a \czmrev{p-size} normalized vector that is used to give more relevance to the opinions of those judges with high weights.
		
\item	
	\czmrev{The} item $I_r$ is assessed according to $C=~\{C_1,\dots,C_q\}$, \czmrev{which is} a set \czmrev{composed by the following} $q=4$ linguistic criteria:
   \begin{itemize}
	\item 
		\emph{Clarity}. \czmrev{It measures the quality of being clear, coherent and intelligible.}
	\item 
		\emph{Writing}. It measures \czmrev{the writing proficiency, \textit{i.e.},} the degree of proofreading in writing.
	\item 
		\emph{Presence}. It measures the pertinence of the item into its dimension. Sometimes the item is well formed but it is placed in the wrong dimension.
	\item 
		\emph{Answering scale}. \czmrev{It measures} the rightness of the answering scale \czmrev{according} to the wording of the item.
	\end{itemize}

\item 
	\czmrev{The} item $I_r$ is also assessed according to a numerical property which characterizes the global \emph{relevance} of the item. \czmrev{It} represents the importance or utility of the item in the questionnaire for the given research hypothesis. Expert $J_i$ rates this property with a number $w_i^r \in [0,1]$ to be used as the item weight in the processes of computing the linguistic result. We note \czmrev{by} $R$ the array of values given by each expert. 
\end{itemize}

The 2-Tuple Fuzzy Linguistic Delphi method is an iterative and dynamic process aimed at achieving a high degree of agreement before making the decision that solves. It uses linguistic and numerical data which is managed as individual assessments with respect to item $I_r$. \czmrev{We compute the following for each item:} (1) the collective group opinion $Y^r$ by the aggregation of all individual opinions of experts with respect to every criteria; (2) the consensus index $CI_r$ by considering all the criteria; and (3) the reliance index $RI_r$ to determine if the item is valid. The previous procedure is repeated for all the elements of $I$, resulting in a new information for the overall questionnaire, that we call Questionnaire Score $QS \in S^7$. 

\begin{figure}[h!]
\begin{center}
\includegraphics[width=13.66cm,height=4cm]{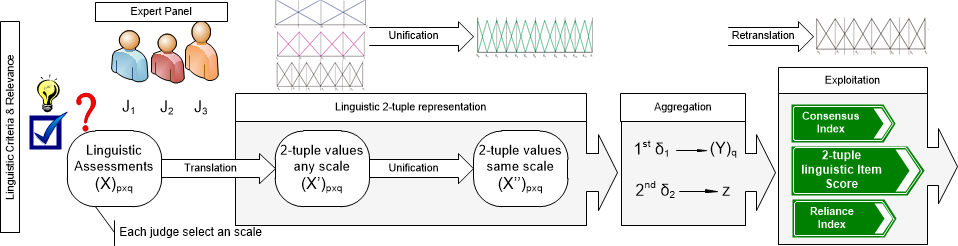}
\end{center}
\caption{The solution of a Multi Expert Multi Criteria Linguistic Decision Making through successive phases is the qualification of a item of the questionnaire.}
\label{fig:single}
\end{figure}

Traditionally, the selection process for reaching a solution for a Linguistic Decision Making problem after the definition of the problem, perform two main phases~\cite{ZUHEROS202122}: (1) aggregation, in which experts opinions are combined by using an aggregation operator, and (2) exploitation, that uses a selection criterion to obtain an alternative or a subset of alternatives as the solution to the problem. However, our proposal does not deal with different alternatives and it handles flexible ways of providing linguistic information. \czmrev{Thus}, we extend the classic processes to solve a Multi-Expert Multi-Criteria Linguistic Decision Making problem \czmrev{as} shown in Figure~\ref{fig:single}.

The following computational processes are better detailed:
\begin{itemize}

	\item \emph{Gathering phase}. 
An opinion is a single label represented with the 2-tuple fuzzy linguistic model according to an extended linguistic hierarchy. A linguistic information $x_{ij}^r \in S^{n(t)}$ is given by judge $J_i$ regarding criterion $C_j$ \czmrev{when evaluating the} item $I_r$ \czm{(see Table \ref{tab:sampleMatrix})}. \czmrev{The linguistic assessment matrix associated to the item $I_r$ is denoted by} $X_r~=~(x_{ij}^r)_{p \times q}$. Experts also express \czmrev{their opinions} regarding the relevance of $I_r$ with numerical values. 

\begin{table}[ht!]
  \centering
  $I_r \longrightarrow$
  \begin{tabular}{c|cccc|c} 
	\hline \hline 
   	 &$C_1$ & $C_2$ &$C_3$ &$C_4$ & $R$\\
	\hline
	\hline $J_1$ 	& $x_{11}^{r}$ 	& $x_{12}^{r}$	& $\dots$	 & $x_{1q}^{r}$ 	& $w_1^r$ \\
	\hline $\vdots$ 	& $\vdots$ 	& $\vdots$	& $\vdots$ & $\vdots$  	& $\vdots$ \\
	\hline $J_p$ 	& $x_{p1}^{r}$ 	& $x_{p2}^{r}$ 	& $\dots$	 & $x_{pq}^{r}$	& $w_p^r$ \\
	\hline \hline 
	\end{tabular} 
  \caption{Full assessment matrix for item $I_r$.}
  \label{tab:sampleMatrix}
\end{table}

	\item \emph{2-tuple transformation phase}. 
Linguistic values are represented using the 2-tuple fuzzy linguistic computational approach, so \czmrev{labels $x_{ij}^r$ are} translated to $(x_{ij}^r,0)$. 
\czmrev{We} note \czmrev{by} $X'_r~=~(x_{ij}^{'r})_{p \times q}$ the matrix of linguistic 2-tuples values.  

	\item \emph{Unification phase}. 
This phase unifies linguistic data by expressing all of them in the same linguistic term set $S^{n(t^*)}$, which is the one with the highest cardinal. \czmrev{The original data are provided into $ELH = S^3 \cup S^5 \cup S^7$, so we transform them into $S^{n(t^*)}$ = $S^{13}$ (see Section \ref{sec2new5}) by means of the function $TF_{t^*}^t$ (see Eq. (\ref{eq:transFunct}))}. We denote \czmrev{by} $X^{''}_r~=~(x_{ij}^{''r})_{p \times q}$ the assessment matrix with \czmrev{the unified} 2-tuple linguistic terms in $S^{13}$.

\begin{example}\label{ex:1}
Opinions over an alternative regarding a criterion are elicited according to different term sets at $n(1), n(2), n(3)$. Unification is resolved as follows:
$$
\begin{array}{l}
	n(1)=3\;;(s_{1}^{3}, 0) \Rightarrow TF_{13}^3 = \Delta(\frac{1 \cdot 12}{2}) = (s_{6}^{13},0)\\
	n(2)=5\;;(s_{3}^{5}, 0) \Rightarrow TF_{13}^5 = \Delta(\frac{3 \cdot 12}{4}) = (s_{9}^{13},0)\\
	n(3)=7\;;(s_{4}^{7}, 0) \Rightarrow TF_{13}^7 = \Delta(\frac{4 \cdot 12}{6}) = (s_{8}^{13},0)\\
\end{array}
$$
\end{example}

	\item \emph{Double aggregation phase}. 
\czmrev{We conduct a first aggregation that combines the opinions to obtain an evaluation of each item regarding each criterion.}
Aggregation is carried out \czmrev{through} the arithmetic weighed extended mean operator $\bar{x}^e(W_{D_m})$ \czmrev{(see Eq.~(\ref{eq:avgTuples})), considering} the expert weights $W_{D_m}$ with respect item $I_r \in D_m$ and for each judge $J_i \in J$. For simplicity in the notation, we \czmrev{set} $\delta_1=\bar{x}^e(W_{D_m})$. \czmrev{Thus, we} apply $\delta_1$ to the linguistic information $X^{''}_r$ \czmrev{getting the q-size} vector \czmrev{$Y_r~=y_{j}^{r}$}
. We also compute the item average relevance $W^r = \frac{1}{p}\sum_{i=1}^p\;w_i^r \cdot w_{iD_m}$\asocrev{, $w_{iD_m} \in W_{D_m}$}. \czmrev{Table \ref{tab:vectorNotation} shows the data format of this first aggregation.}

\begin{table}[ht]
  \centering
  \begin{tabular}{c|cccc|c} 
	\hline \hline 
   	 &$C_1$ & $C_2$ &$C_3$ &$C_4$ &$R$ \\
	\hline
	\hline $\delta_1(X^{''}_r)=Y_r$ 	& $y_{1}^{r}$ 	& $y_{2}^{r}$	& $\dots$	 & $y_{q}^{r}$ & $W^r$ \\
	\hline \hline 
  \end{tabular} 
  \caption{Vector notation for the first aggregation.}
  \label{tab:vectorNotation}
\end{table}

The next step considers a second aggregation over the 2-tuple linguistic values $y_j^r$. In this way, we compute the collective overall opinion for item $I_r$ by aggregating criteria values with the aggregation operator $\bar{x}^e(V)=\delta_2$ instantiated with a vector of uniform weights ($V=\{v_j=1/q \;\;|\;\; j=1,\dots,q\}$). 

\begin{equation}\label{eq:collective}
	\delta_2(Y_r)=Z_r~=~(s_z^r,\alpha_z^r) \;\;\;with\;\;\;s_z^r \in S^{n(t^*)}
\end{equation}

	\item \emph{Exploitation phase}. 
We are able to provide two types of results:
    \begin{itemize}
	\item 
	Individual results for item $I_r$. The value $(s_z^r,\alpha_z^r)$ from Eq.~(\ref{eq:collective}) is the main output to the expert panel. Nevertheless, we also have to compute \textcolor{black}{the Consensus Index ($CI$) and the Reliance Index ($RI$)} for the knowledge of the moderator. Boolean values \textcolor{black}{the Consensus Status ($CS$) and Reliance Status ($RS$)} could be shared with the expert panel to better understand the circumstances for a new iteration. For the sake of understanding, given that experts $J$ expressed themselves over $S^3$ or $S^5$ or $S^7$, we apply a re-translation from level $t^*$ to level $t=h=3$. The linguistic output is known as the Item Score $IS_r$ with $s_r \in \{Dreadful, Incorrect, Moderate, Correct, Very\,correct,Excellent\}$.
\begin{equation}\label{eq:ToS7}	
	IS_r~=~TF_{t^*}^{3}(s_z^r,\alpha_z^r)~=~(s_r,\alpha_r)
\end{equation}
	\item 
	Collective results for questionnaire $Q$. A questionnaire is in fact a set of $n$ items, each one with a linguistic score $IS_r$ and a collective relevance opinion $W^r$. \czmrev{The} 2-Tuple Fuzzy Linguistic Delphi method \czmrev{computes global scores for $Q$ as 2-tuple linguistic values in $S^7$ by using the} relevance values as weights in a third aggregation step $\delta_3=\bar{x}^e(CW)$ with $CW = \frac{1}{n} \sum_{r=1}^n W^r$\czmrev{, getting}:
	\begin{itemize}
		\item Collective Clarity: $CC = \delta_3(Y_{C_1})$ with $Y_{C_1} = \{y_1^1, \dots,y_1^n\}$.
		\item Collective Writing: $CW = \delta_3(Y_{C_2})$ with $Y_{C_2} = \{y_2^1, \dots,y_2^n\}$.
		\item Collective Presence: $CP = \delta_3(Y_{C_3})$ with $Y_{C_3} = \{y_3^1, \dots,y_3^n\}$.
		\item Collective Answering Scale: $CAS = \delta_3(Y_{C_4})$ with $Y_{C_4} = \{y_4^1, \dots,y_4^n\}$.
		\item Questionnaire Score $QS = \delta_3(CIS)$ with $CIS = \{IS_1, \dots,IS_n\}$.
	\end{itemize}	
	 
   \end{itemize}

\end{itemize}

\subsection{Validation by \czm{a Consensus Model}}\label{section35}

The consensus process tries to achieve the maximum degree of consensus possible among the opinions of individuals or experts. The degree of consensus is calculated at each \czmrev{iteration. The} questionnaire is positively tested \czmrev{when the consensus} grade is satisfactory. Conversely, if the degree of consensus is not satisfactory, then individuals or experts are encouraged to modify their views in order to increase proximity in their approaches. In this way, we set a \czmrev{ dynamic and iterative} Decision Making process in which experts change their opinions until their approaches to the solution are sufficiently close, at which point, consensus is reached. 

\czmrev{We} need to measure the difference between individuals and collective opinions \czmrev{to} compute the degree of consensus\czmrev{,} which is in fact a measure of error. \czmrev{It} is desirable to get differences values close to zero, meaning that \czmrev{expert} opinions are similar. \czmrev{There are} $p$ judges \czmrev{that provide} $n$ linguistic decision matrices $X'_r$ \czmrev{(see Section~\ref{section34})}. We also have $n$ 2-tuple linguistic collective opinion $Y^r$ with respect to a criteria set of $q$ elements. Finally, for each item, the 2-tuple linguistic output value is $(s_z^r, \alpha_z^r)$. Also per item, we compute a separation measure $\rho \in [0,\infty)$ for each judge $J_i$:

\begin{equation}\label{eq:separation}
	\rho_i = \sqrt{ \sum_{j=1}^q \left( \Delta^{-1}(x'_{ij}) - \Delta^{-1}(y_{j}) \right) ^2}\;,\;\;i~=~1,\dots,p
\end{equation}

\czmrev{High $\rho$ values indicates to the corresponding experts that, in general, their opinions are not very similar to those of the collective.}

\theoremstyle{definition}\label{def:consensus}
\begin{definition}
Let Consensus Index $CI_r \in [0,1]$ be \czmrev{the consensus} between experts regarding item $I_r$. We consider that the information collected from the judges could be influenced by vectors of normalized weights $\{W_{D_1},\dots,W_{D_l}\}$ that represent the expertise degrees defined for each dimension of the questionnaire, were $v_i \in W_{D_m}$ if $I_r \in D_m$. Let $CS_r$ be a boolean value that takes \emph{true} if there is consensus, that is when $CI_r \ge 0.5$ or \emph{false} in other case. According to these assumptions, the consensus index is defined as:

\begin{equation}\label{eq:CI}
	CI_r~=~ 1 - \frac{\sum_{i=1}^p \rho_i \cdot v_i}{\delta_{t^*}}\\
\end{equation}
\end{definition}

In our opinion, consensus processes need to be flexible and adjustable by the moderator. 
\czmrev{Thus, we} use a parameter called \emph{satisfactory reliance level} $\epsilon \in [0,1]$ \czmrev{to} determine \czmrev{the} consensus \czmrev{that} can be reached in \czmrev{certain} number of iterations. \czmrev{When $\epsilon$ approaches to one, it becomes increasingly difficult for experts with high $\rho_i$ values to narrow the gap with the group. This} parameter represents \czmrev{whether} the solution is acceptable to the moderator. 

\begin{definition}\label{def:reliance}
We define the Reliance Index $RI_r \in [0,1]$ of an item $I_r$ \czmrev{by}:

\begin{equation}\label{eq:RI}
	RI_r~=~ \sum_{j=1}^q u_j \;\;\;\; where\;\;u_j =
	\left\lbrace 
  	\begin{array}{ll}
     		1/q & if~ \Delta^{-1}(y_j) \ge \delta_{t^*} \epsilon\;,\\
     		0 & else.
  	\end{array} \right. 
\end{equation}
\end{definition}

In this sense, $RS_r$ is the boolean value that takes \emph{true} when $RI_r \ge \epsilon$ or \emph{false} in other case. Note that in assessment phase, Computing with Words processes are done at level $t^*$. 

\begin{example}\label{ex:2}
In this example moderator sets $\epsilon=0.6$. Consider that opinions regarding $I_1$ from $J=\{J_1,J_2,J_3\}$ are the same that in Example~\ref{ex:1}, thus $X' = \{(s_6^{13},0), (s_9^{13},0), (s_8^{13},0)\}$ with $W_{D_1}=(0.2,0.6,0.2)^T$. For simplicity, we assume single criteria ($q=1$) and thus $(s_z,\alpha_z)=(s_8^{13}, 0.2)$. 

By using Eq.(\ref{eq:separation}) we have $\rho = \{ 2.2, 0.8, 0.2\}$ that reflects that $J_1$ has the most distant opinion from the solution, nevertheless $v_1=0.2$ is low, and total consensus is positive with $CI_1=0.92$ by applying Eq.(\ref{eq:CI}). The previous values set $CS_1=true$ and $RS_1=true$ because $RI_1=1$. 

Yet, if we change the model parameter to $\epsilon=0.8$ the overall situation changes, given that the inequality $\Delta^{-1}(s_8^{13}, 0.2) \ngeq 9.6$ from Eq.(\ref{eq:RI}). As a result, it sets $RI_1=0$ and $RS_1=false$. In this situation, the moderator needs another round of assessments to improve consensus and reliance levels.

\end{example}

\section{A web tool based DSS to apply the 2-Tuple Fuzzy Linguistic Delphi method}\label{section4}
Making decisions is a mentally demanding act\czmrev{, so DSS tools are designed to assist this task.} This work contributes with a web tool DSS that implements the 2-Tuple Fuzzy Linguistic Delphi method presented in Section~\ref{section3}. In the following sections, we describe the requirements of this software \czmrev{highlighting its most outstanding features (see Section~\ref{section41}) and} how it accepts the information through user-supplied input files (see Section~\ref{section42}).

\subsection{The 2-tuple-fuzzy-delphi DSS}\label{section41}
\czmrev{There are some online solutions to the Delphi method including both free, such as Delphi2\footnote{Delphi2 \url{http://armstrong.wharton.upenn.edu/delphi2/}}, and commercial software, such as Mesydel\footnote{Mesydel \url{http://www.spiral.ulg.ac.be/en/tools/online-delphi-mesydel/}} and Surveylet\footnote{Surveylet \url{https://calibrum.com}}.} These tools are not licensed for adaptation or modification, so it is difficult to put in practice suitable linguistic representation models or solution schemes of Computing with Words. \czmrev{It} indicates the Delphi method is active in the research community and there is an opportunity to assist the iterative processes reducing the cost of applying the method. 

\czmrev{We propose the \textit{2-tuple-fuzzy-delphi}\footnote{\czmrev{The 2-tuple-fuzzy-delphi DSS software} is available at \href{https://sci2s.ugr.es/2tuple-fuzzy-delphi}{https://sci2s.ugr.es/2tuple-fuzzy-delphi} for public use. Its code source is available under GNU GPL v3 license at GitHub repository \href{https://github.com/ari-dasci/S-2tuple-fuzzy-delphi}{https://github.com/ari-dasci/S-2tuple-fuzzy-delphi}.} DSS as an online tool that guides the moderator in the task of reaching a consensual questionnaire. It applies the 2-Tuple Fuzzy Linguistic Delphi method. Particularly, at} each iteration, \czmrev{the moderator} imports the original assessments from the panel of experts $J$ to visualize the individual and collective linguistic scores extracted from the solution of the Multi-Expert Multi-Criteria Linguistic Decision Making problem and the overall consensus. The questionnaire can be used for piloting \czmrev{when the reliance and consensus levels are satisfactory}. Otherwise, the tool can be used to provide feedback to the expert panel.

\begin{figure}[h!]
\begin{center}
\includegraphics[width=.9\textwidth,height=5.5cm]{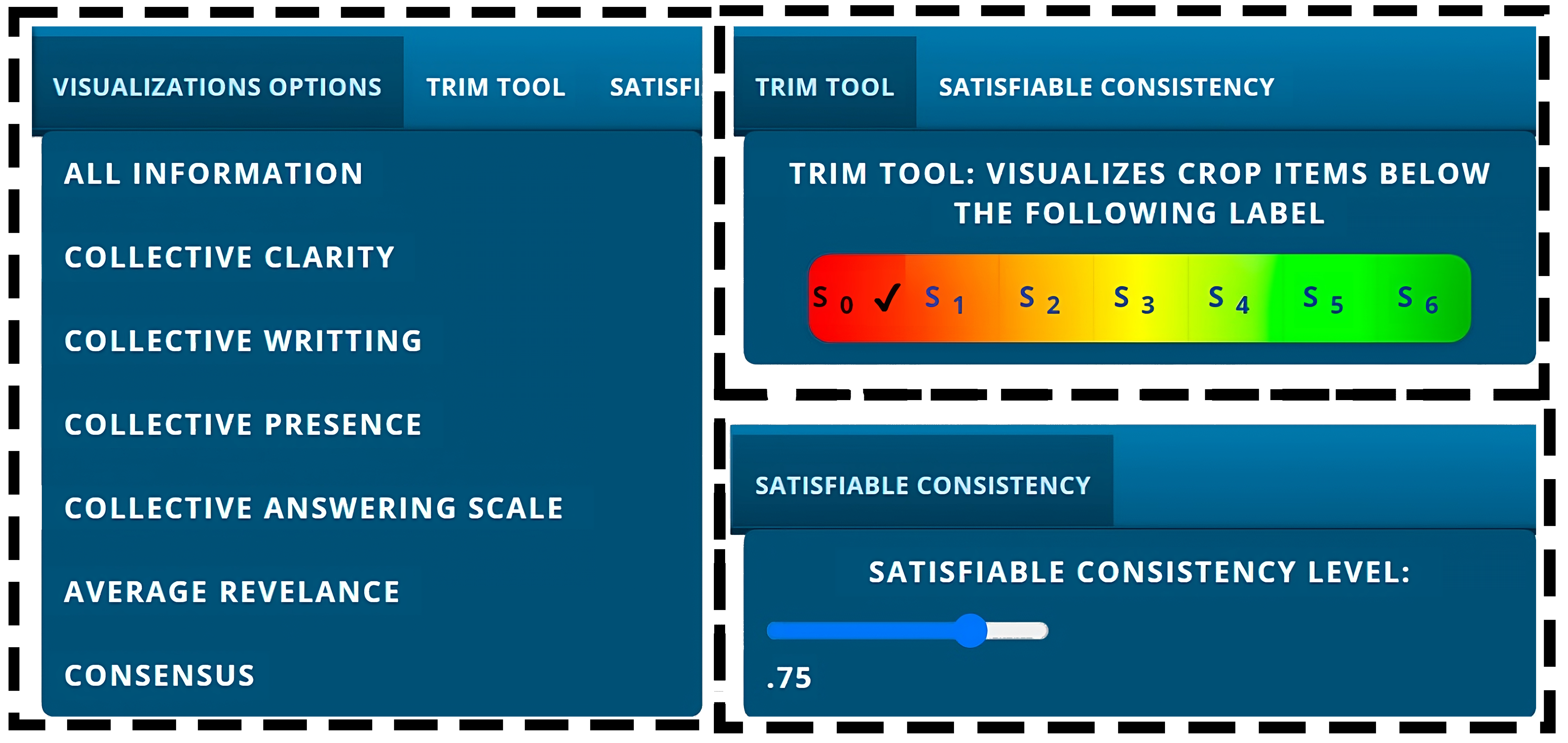}
\end{center}
\caption{Main features of the DSS web tool.}
\label{fig:features}
\end{figure}

\czmrev{The} key features of \textit{2-tuple-fuzzy-delphi} DSS are shown in Figure \ref{fig:features} and described bellow:
\begin{itemize}
	\item \emph{Filtering:} \czmrev{The user can visualize the data, even restricted to particular columns, by selecting different filtering options: all information, collective clarity, collective writing, collective presence, collective answering scale, average relevance, and consensus.}
	\item \emph{Trimming:} \czmrev{The expert panel may advice} a reduction in the number of items. \czmrev{The DSS provides a trim tool that assists to solve the question:} \emph{which elements should be removed to address my desirable number of items?} \czmrev{The trim operator considers the linguistic term set $S^7$ as input scale. By default, it is set to $s_0$ meaning zero trimmed items. By increasing the linguistic term, some items are hidden and a label reports the number of trimmed items.} 
	\item \emph{Data simulation:} \czmrev{The user can adjust, into the satisfiable consistency navigation bar, the model solution by using a slider parameter that allows to set different acceptable levels of consensus.}
	\item \emph{Searching:} There is a text searching tool which makes it easy \czmrev{for the user} to locate an item and focus on its scores.
	\item \emph{Sorting:} \czmrev{The user can sort the tabular data in increasing order from A to Z and in decreasing order from Z to A. }
\end{itemize}

\subsection{The 2-tuple-fuzzy-delphi DSS input file format}\label{section42}
\czmrev{The data input files to be used in each iteration of the \textit{2-tuple-fuzzy-delphi} DSS tool are in CSV format because of two reasons: (1) it is common to conduct questionnaires through online surveys supported by some well known online services, such as Google Forms\footnote{Google Forms \url{https://www.google.es/intl/en/forms/about/}}, Monkey Survey\footnote{Monkey Survey \url{https://www.surveymonkey.com}} or Lime Survey\footnote{Lime Survey \url{https://www.limesurvey.org}}, which allows to export the answers in CSV files, and (2) a CSV} file is easily get from spreadsheets desktop solutions when the researcher does not use the previous services. \czmrev{This way, the proposed tool} is able to nourish data from external sources that are highly available.

\czmrev{We define the input format guided by how Google Forms exports its content data as spreadsheets. Google Forms stores responses in a \emph{.gsheet} file that can be downloaded in OpenOffice or Microsoft Office format. In any case, particular sheets can be exported to CSV file format individually. The moderator starts using the \emph{2-tuple-fuzzy-delphi} DSS after the first round is complete.} Generally for each round we can use up to three type of sheets. We suggest the following wording to be used as the sheet names, where $X$ represents the number of the current iteration. The following is a description of the content to be stored on each sheet:

\begin{itemize}
\item 
	\emph{Round$X$Description}. \czmrev{It contains} the text description per item. This import is not mandatory as a generic text would be used in case of absence. \czmrev{The first row could be} the header name\czmrev{, in which case} $n$ is the number of lines read minus 1. Content type in this case is: \emph{description}. 

\item 
	\emph{Round$X$Dimensions}. \czmrev{It} associates each judge's expertise with questionnaires dimensions \czmrev{as well as} items ranges with dimensions. Table~\ref{tab:dimMatrix} presents its structure. The number of lines (minus one if headers are enabled) is $l$, the number of dimensions. This import is not mandatory as uniform weights would be used in case of absence. Content type in this case is: \emph{dimensions}. 
	
\item
	\emph{Round$X$Responses}. It contains the Multi-Expert Multi-Criteria Linguistic Decision Making problem data, so it is mandatory. \czmrev{Table~\ref{tab:sampleMatrixv2} presents its structure.} According to the number of rows an columns parsed, we compute the number of experts $p$ and the number of items $n$ respectively. Content type in this case is: \emph{responses}. 
\end{itemize}

\begin{table}[ht]
  \centering
  \begin{tabular}{c|cc|cccc} 
	\hline \hline 
	Dimension & Begin & End & $J_1$ & $J_2$ & \dots & $J_p$\\
	\hline \hline 
	$D_1$ & $I_1$ & $I_i$ &  $w_{1D_1}$  &  $w_{2D_1}$ & $\dots$ & $w_{pD_1}$\\
	\hline 
	$D_2$ & $I_{i+1}$ & $I_j$ & $w_{1D_2}$ & $w_{2D_2}$ & $\dots$ &  $w_{pD_2}$\\
	\hline $\vdots$	& $\vdots$ & $\vdots$  & $\vdots$ & $\vdots$ & $\dots$ & $\vdots$ \\
	\hline 
	$D_l$ & $I_v$  & $I_n$ & $w_{1D_l}$ & $w_{2D_l}$ & $\dots$ &  $w_{pD_l}$\\	

	\hline \hline 
  \end{tabular} 
  \caption{Structure of the data grid: subdivision of items into dimensions and expert weights per dimensions of the questionnaire. Header names are optional.}
  \label{tab:dimMatrix}
\end{table}

\begin{table}[ht]
  \centering
  \begin{tabular}{cc|ccccc|c|ccccc} 
	\hline \hline 
   	 Judge & Level &$C_1$ & $C_2$ &$C_3$ &$C_4$ & $R$  & $\dots$ &$C_1$ & $C_2$ &$C_3$ &$C_4$ & $R$\\
	\hline
	\hline $J_1$ & $n(t)_{J_1}$ & $x_{11}^{1}$ & $x_{12}^{1}$	& $\dots$	 & $x_{1q}^{1}$ 	& $w_1^1$ 
		& $\dots$ & $x_{11}^{n}$ & $x_{12}^{n}$	& $\dots$	 & $x_{1q}^{n}$ 	& $w_1^n$ 
	\\
	\hline $J_2$ & $n(t)_{J_2}$ & $x_{21}^{1}$ 	& $x_{12}^{1}$	& $\dots$	 & $x_{2q}^{1}$ 	& $w_2^1$ 
		& $\dots$ & $x_{21}^{n}$ & $x_{22}^{n}$	& $\dots$	 & $x_{2q}^{n}$ 	& $w_2^n$ 	
	\\	
	\hline $\vdots$ 	& $\vdots$ 	& $\vdots$	& $\vdots$ & $\dots$  	& $\vdots$ & $\vdots$  
		& $\vdots$ 	& $\vdots$	& $\vdots$ & $\dots$  	& $\vdots$ & $\vdots$ 
	\\
	\hline $J_p$ & $n(t)_{J_p}$ 	& $x_{p1}^{1}$ 	& $x_{p2}^{1}$ 	& $\dots$	 & $x_{pq}^{1}$	& $w_p^1$ 
		& $\dots$ & $x_{p1}^{n}$ & $x_{p2}^{n}$	& $\dots$	 & $x_{pq}^{n}$ 	& $w_p^n$ 
	\\
	\hline \hline 
  \end{tabular} 
  \caption{Structure for $Q$ data grid is similar to Google Form spreadsheets responses grid. Header names are optional.}
  \label{tab:sampleMatrixv2}
\end{table}

\begin{figure}[h]
\begin{center}
\includegraphics[width=13.66cm,height=6.9cm]{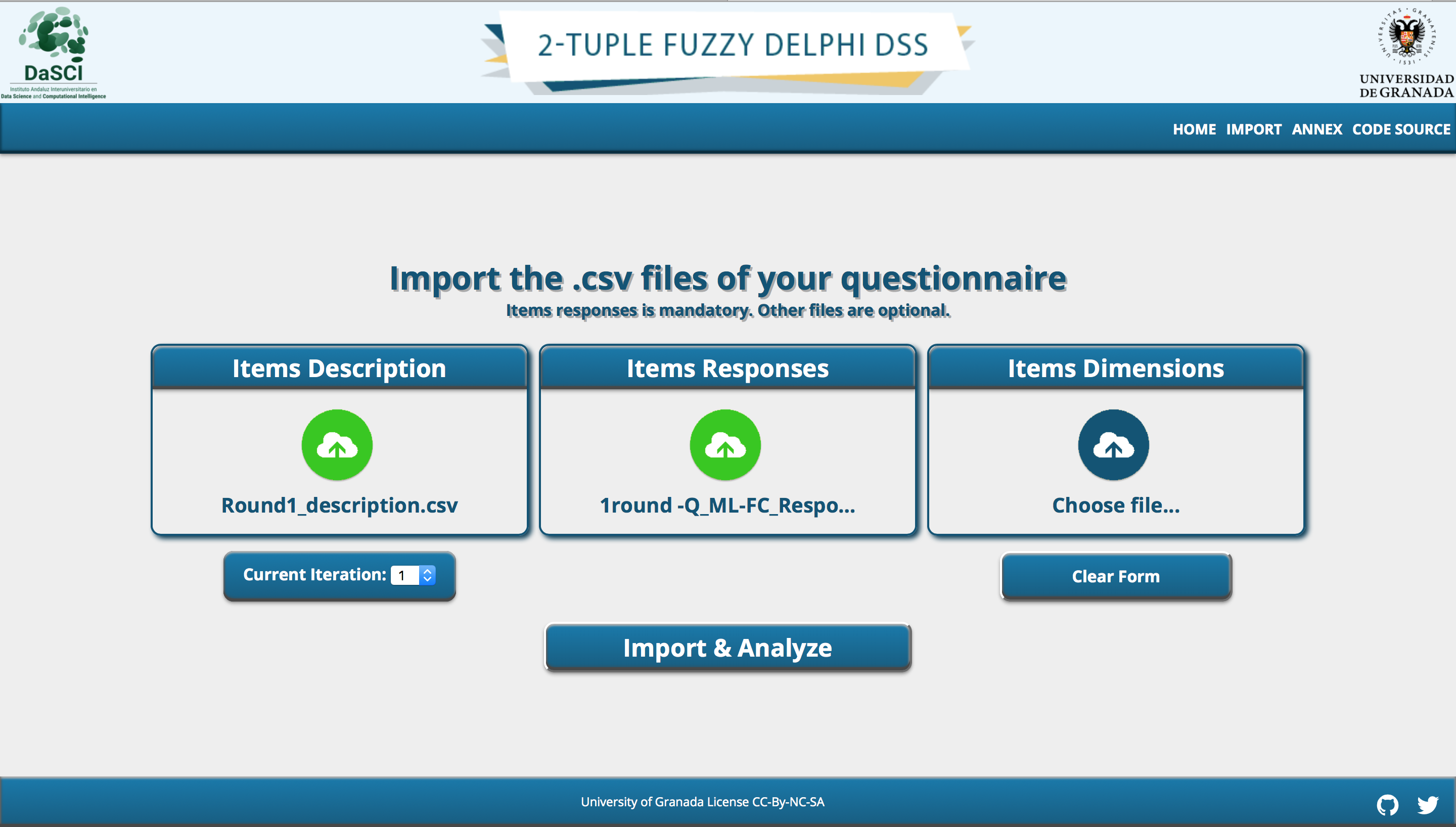}
\end{center}
\caption{We can import separately the description of the questionnaire and the assessments of the expert panel for each round.}
\label{fig:import}
\end{figure}

\czmrev{Figure~\ref{fig:import} shows how each sheet can be exported separately as a CSV file and then imported into the \emph{2-tuple-fuzzy-delphi} DSS online tool. We require the number of the current round, the content type, and the file data path. }

\section{Case Study: content validity of a questionnaire for b-learning}\label{section5}

We have conducted a \czmrev{b-learning} experience in education using the Flipped Classroom and the m-Learning methodologies in combination. \czmrev{We want to evaluate the students satisfaction with the experience in the course \emph{Fundamentals of Software} by means of a questionnaire with $45$ items (see Section~\ref{section2}). In this section, we apply the 2-Tuple Fuzzy Linguistic Delphi method to ensure the content validity of the questionnaire.} We configured a committee of $p=9$ experts. Table~\ref{tab:dimBL} describes the expertise of each judge according the structure of the questionnaire and how items are grouped in $7$ dimensions. This is the content of the \emph{dimensions} sheet. 
\begin{table}[ht]
   \centering
   \resizebox{14cm}{!} {
	\begin{tabular}{c|cc|ccccccccc}
		\hline \hline
		Dimension & Begin & End & $J_1$ & $J_2$ & $J_3$ & $J_4$ & $J_5$ & $J_6$ & $J_7$ & $J_8$ & $J_9$ \\ 
		\hline \hline
		$D_1$ & $1$ & $8$ & $0.118$ & $0.093$ & $0.087$ & $0.124$ & $0.112$ & $0.124$ & $0.112$ & $0.124$ & $0.106$ \\ 
		\hline 
		$D_2$ & $9$ & $14$ & $0.125$ & $0.094$ & $0.088$ & $0.119$ & $0.113$ & $0.113$ & $0.113$ & $0.125$ & $0.113$ \\ 
		\hline 
		$D_3$ & $15$ & $21$ & $0.101$ & $0.094$ & $0.094$ & $0.126$ & $0.113$ & $0.126$ & $0.113$ & $0.126$ & $0.107$ \\ 
		\hline 
		$D_4$ & $22$ & $28$ & $0.121$ & $0.096$ & $0.089$ & $0.127$ & $0.115$ & $0.127$ & $0.115$ & $0.102$ & $0.108$ \\ 
		\hline 
		$D_5$ & $29$ & $35$ & $0.133$ & $0.100$ & $0.093$ & $0.080$ & $0.120$ & $0.133$ & $0.120$ & $0.107$ & $0.113$ \\ 
		\hline 
		$D_6$ & $36$ & $41$ & $0.123$ & $0.097$ & $0.091$ & $0.130$ & $0.117$ & $0.110$ & $0.117$ & $0.104$ & $0.110$ \\ 
		\hline 
		$D_7$ & $42$ & $45$ & $0.116$ & $0.098$ & $0.091$ & $0.122$ & $0.110$ & $0.110$ & $0.122$ & $0.110$ & $0.122$ \\ 
		\hline \hline 
	\end{tabular} 
   }
   \caption{Structure of the dimension grid: expert weights regarding dimensions of the questionnaire along with the subdivision of items into dimensions.}
   \label{tab:dimBL}
\end{table}

\czmrev{The consensus phase of the 2-Tuple Fuzzy Linguistic Delphi method required two iterations to solve the present case study.} In the first iteration, we got a questionnaire score $QS_1 = (s_5, -0.226)$ or \emph{Very correct} \czmrev{along with a description of the changes applied to the questionnaire (see Section~\ref{section51}).} In the second iteration we got $QS_2 = (s_6, -0.282)$ or \emph{Excellent} \czmrev{(see Section~\ref{section52}), achieving a consensual version of the questionnaire.} 
\czmrev{The \ref{annexA} presents the items of the final version of the questionnaire.}

\subsection{Applying 2-Tuple Fuzzy Linguistic Delphi method: first round}\label{section51}

\czmrev{We} focus on one of the items that, in the first round, \czmrev{is a} point of conflict between the judges \czmrev{in order to} describe the application of the method in a simplified way. This is item $I_{27}$ \czm{whose Spanish text can be translated as \textit{``I consider that I have achieved the objectives of the course.'' Scale to be used: Type B.}} In this case, experts $J_1-J_3$ selected the linguistic term set $S^3$, expert $J_9$ set $S^5$ and experts $J_4 - J_8$ decided to perform the evaluations using $S^7$. According to Section~\ref{section34}, \czmrev{we undertake} the following computational processes\czmrev{:}
\begin{itemize}
	\item \emph{Gathering phase}. 
	Table~\ref{tab:dataI27} shows the original assessments in consideration of ELH $h=3$, the matrix $(X_{27})_{9\times4}$. This information is stored in the \emph{responses} sheet. 
	
	\item \emph{2-tuple transformation phase}. 
	Linguistics values with respect to criterion $C_1$ to criterion $C_4$ are transformed into 2-tuples linguistic values by the application of Eq.~(\ref{eq:2Tuple}).
	
	\item \emph{Unification phase}. 
	A\czmrev{n} unified vision of data is achieved through the application of the transformation function given at Eq.~(\ref{eq:transFunct}). In the particular case of $I_{27}$, Table~\ref{tab:unifiedI27} shows intermediate results $(X_{27}^{''})_{9\times4}$ of performing both transformational steps.
	
	\item \emph{Double aggregation phase}. We perform two rounds of aggregation with $\delta_1$ and $\delta_2$. According to expression Eq.~(\ref{eq:collective}) we obtain $(Y_{27})_{4}$, $W^{27}$ and $Z_{27}=(s_9, 0.263)$.
	
	\item \emph{Exploitation phase}. \czmrev{It outcomes} the re-translation of the linguistic solution to the 2-tuple-fuzzy-linguistic scale, $IS_{27}= (s_5, -0.369) \in S^7$. \czmrev{It also provides} the evaluation of the consensus degree obtained for this item. By using Eq.(\ref{eq:separation}) we have the following vector:
	$$
		\rho = \{7.679, 6.407, 4.482, 6.368, 5.858, 6.407, 1.995, 6.088, 9.180 \}
	$$
	Using $\rho$ we identify judges $J_1$ and $J_9$ as distant from the consensus. They may affect the consensus level  $CS_{27}=false$ because the consensus index obtained with Eq.(\ref{eq:CI}) is $CI_{27}=0.493$. With $\epsilon=0.75$ we get $RS_{27}=false$ and $RI_{27}=0.5$. 
	
\end{itemize}

\begin{table}[ht]
  \centering
	\begin{tabular}{c|c|c|c|c|c}
		\hline \hline
		\multicolumn{6}{c}{First round for $I_{27}$} \\ 
		\hline \hline
		Judge & Clarity & Writing & Presence & A.Scale & Relevance\\ 
		\hline 
		$J_1$ & $s_{2}^{3}$ & $s_{0}^{3}$ & $s_{2}^{3}$  & $s_{1}^{3}$ & $1.00$\\ 
		\hline 
		$J_2$ & $s_{2}^{3}$ & $s_{2}^{3}$ & $s_{2}^{3}$ & $s_{2}^{3}$ & $1.00$\\ 
		\hline 
		$J_3$ & $s_{2}^{3}$ & $s_{1}^{3}$ & $s_{2}^{3}$ & $s_{2}^{3}$ & $1.00$\\ 
		\hline 
		$J_4$ & $s_{5}^{7}$ & $s_{6}^{7}$ & $s_{6}^{7}$ & $s_{6}^{7}$ & $1.00$\\ 	
		\hline 
		$J_5$ & $s_{4}^{7}$ & $s_{3}^{7}$ & $s_{4}^{7}$ & $s_{2}^{7}$ & $0.90$\\ 
		\hline 
		$J_6$  & $s_{6}^{7}$ & $s_{6}^{7}$ & $s_{6}^{7}$ & $s_{6}^{7}$ & $1.00$\\ 
		\hline 
		$J_7$ & $s_{6}^{7}$ & $s_{3}^{7}$ & $s_{6}^{7}$ & $s_{4}^{7}$ & $1.00$\\ 
		\hline 
		$J_8$  & $s_{4}^{7}$ & $s_{4}^{7}$ & $s_{3}^{7}$ & $s_{3}^{7}$ & $1.00$\\ 
		\hline 
		$J_9$ & $s_{4}^{5}$ & $s_{1}^{5}$ & $s_{4}^{5}$ & $s_{0}^{5}$ & $0.99$\\ 
		\hline \hline 
	\end{tabular} 
  \caption{Gathered opinion regarding item $I_{27}$ considering the four linguistic criteria and the pertinence.}
  \label{tab:dataI27}
\end{table}

    Now the moderator analyzes with detail the data obtained with the use of \czmrev{the} \emph{2-tuple-fuzzy-delphi} \czmrev{DSS} tool in order to undertake modifications in $I_r$ \asocrev{(such as in the case of $I_{27}$)} and later by extension, into the full questionnaire. According to all the experts, the criteria are valuated as:
	
	$$
		Y_{27} = \{ (s_{11}, -0.122), (s_7, 0.254), (s_{11}, -0.072), (s_8, -0.014)\}
	$$

The item $I_{27}$ is quite well valued considering criterion $C_1$ and $C_3$, and the nine experts have considered that the relevance of this question in $Q$ is $W^{27}=0.987$. The value of $RI_{27}$ also tells moderator that is a good item, though not perfect. Thus\czmrev{,} $I_{27}$ is not rejected but modified. Considering that $C_2$ represents an evaluation about the writing, the text of this item is changed \czm{so it can be translated as \textit{``I am satisfied with the achievement of the objectives of the course.'' Scale to be used: Type B. }}
This new description is updated in the instance of the questionnaire in Google Forms, and also in the content of the \emph{description} sheet.

\begin{table}[t]
	\centering
	\begin{tabular}{c|c|c|c|c}
		\hline \hline 
		Judge & Clarity & Writing & Presence & Scale \\ 
		\hline \hline
		$J_1$ &($s_{12}^{13}$,0) & ($s_{0}^{13}$,0) & ($s_{12}^{13}$,0) & ($s_{6}^{13}$,0) \\ 
		\hline 
		$J_2$ & ($s_{12}^{13}$,0) & ($s_{12}^{13}$,0) & ($s_{12}^{13}$,0) & ($s_{12}^{13}$,0)\\ 
		\hline 
		$J_3$ & ($s_{12}^{13}$,0) & ($s_{6}^{13}$,0) & ($s_{12}^{13}$,0) & ($s_{12}^{13}$,0) \\ 
		\hline 
		$J_4$ & ($s_{10}^{13}$,0) & ($s_{12}^{13}$,0) & ($s_{12}^{13}$,0) & ($s_{12}^{13}$,0) \\ 
		\hline 
		$J_5$ & ($s_{8}^{13}$,0) & ($s_{6}^{13}$,0) & ($s_{8}^{13}$,0) & ($s_{4}^{13}$,0)\\ 
		\hline 
		$J_6$ & ($s_{12}^{13}$,0) & ($s_{12}^{13}$,0) & ($s_{12}^{13}$,0) & ($s_{12}^{13}$,0) \\ 
		\hline 
		$J_7$ & ($s_{12}^{13}$,0) & ($s_{6}^{13}$,0) & ($s_{12}^{13}$,0) & ($s_{8}^{13}$,0) \\ 
		\hline 
		$J_8$ & ($s_{8}^{13}$,0) & ($s_{8}^{13}$,0) & ($s_{6}^{13}$,0) & ($s_{6}^{13}$,0) \\ 
		\hline 
		$J_9$ & ($s_{12}^{13}$,0) & ($s_{3}^{13}$,0) & ($s_{12}^{13}$,0) & ($s_{0}^{13}$,0) \\ 
		\hline \hline
	\end{tabular} 
	\caption{After the 2-tuple transformation and unification phases, the assessments are prepared for computing with words under the same scale $S^{13}$.}
	\label{tab:unifiedI27}
\end{table}

For the rest of the questionnaire, the expert panel gave several suggestions most of them addressed to grammar (\emph{use of plural and singular must match}), writing issues (were the case of $I_5, I_6, I_7, I_8$) and the answering scale (\emph{the \czmrev{``}satisfied" scale doesn't match my positive impression}). Other comment frequently mentioned was: \emph{It is recommended that the wording of the question be homogeneous with respect to others}. This means that though the expert assess a single item each time, this person maintains an overall record of the questionnaire. Implies also that the last items of a dimension might be penalized in their valuations, not by the item itself (that may be perfectly formed and written), but because homogeneity. Thus, the consistency in the style of writing could be considered as a new  criterion or as part of the instructions given to judges to consider in $C_2$.

In our case, the first round was most oriented to improve the wording, but still an early stage to detect consensus problems. For instance, everyone agreed that $I_{17}$ is not so reliable with $RI_{17}=0.25$. \czmrev{Table~\ref{tab:round1round2} presents the full description of this round.}

\subsection{Applying 2-Tuple Fuzzy Linguistic Delphi method: second round}\label{section52}

The second iteration collects all the assessments given by the judges after receiving the new \czmrev{questionnaire} $Q'$ and a document report with a description of the reliance and consensus status, $RS$ and $CS$ respectively, along with $IS_r$. To describe the second round of the 2-Tuple Fuzzy Linguistic Delphi method, we take up again the valuation of item $I_{27}$, and later we compare the output of the two rounds for the whole questionnaire. The following computational processes are undertaken:

\begin{itemize}
	\item \emph{Gathering phase}. 
	Table~\ref{tab:dataI27round2} shows the original assessments $(X_{27})_{9\times4}$. This time everyone individually selected $S^7$.  
	\item \emph{2-tuple transformation phase}. 
	This step creates matrix $(X'_{27})_{9\times4}$ of 2-tuples linguistic values by the application of Eq.~(\ref{eq:2Tuple}).
	\item \emph{Unification phase}. 
	By the use of ELH aggregation operations happens in level $t^*=4$. After the application of $TF^{n(t)}_{t^*}$ (see Eq.~(\ref{eq:transFunct})), we get 
	$(X''_{27})_{9\times4}$, at it is given in Table~\ref{tab:unifiedI27round2}.
	\item \emph{Double aggregation phase}. Using operator $\delta_1$ we aggregate over the expert opinions, and using operator $\delta_2$ we aggregate over the criteria. We get $W^{27}=0.988$ and:
	$$
		Y _{27}~=~\{(s_{12}, 0), (s_{12}, -0.384) , (s_{12}, -0.254) , (s_{12},-0.217)\} 
	$$
	Now the item is best valorized with regards to the \czmrev{four criteria.} 
	\item \emph{Exploitation phase}. 
	\czmrev{We} re-translate $Z_{27}=(s_12, -0.214)$ to an upper level of the ELH as $TF{t^*}_{n(3)}(Z_{27})=(s_6, 0.107)=IS_{27}$. Again it is a better qualification, but we need to measure if everyone agrees with this result.  By using Eq.(\ref{eq:separation}) we have the following vector:
	$$
		\rho = \{0.254, 1.817, 0.254, 0.254, 0.254, 0.900, 0.254, 0.254, 0.921 \}
	$$
	Previous distant judges $J_1$ and $J_9$ now are close to the group. Only $J_2$ differs low. Using Eq.(\ref{eq:CI}), the consensus index is computed  
	as $CI_{27}=0.907$ (very close to $1$). Applying Eq.(\ref{eq:CI}) we get a reliance index of $RI_{27}=1$\czmrev{, so} both markers are positive, $CS_{27}=true$ and $RS_{27}=true$.
\end{itemize}

Related to the general performance of the questionnaire the previous situation is generalized: item scores are increased, consensus is achieved and reliance is validated. \czmrev{Table~\ref{tab:round1round2} presents the full description of this round.} By comparison of round one and round two we can determine that $Q'$ is a consensual valid questionnaire for data collection regarding constructs: satisfaction in a community of inquiry and virtual communication in a community inquiry for a blended learning experience.

\begin{table}[ht]
	\centering
	\begin{tabular}{c|c|c|c|c|c}
		\hline \hline
		\multicolumn{6}{c}{Second round for $I_{27}$} \\ 
		\hline \hline
		Judge & Clarity & Writing & Presence & A.Scale & Pertinence\\ 
		\hline 
		$J_1$ & $s_{6}^{7}$ & $s_{6}^{7}$ & $s_{6}^{7}$  & $s_{6}^{7}$ & $1.00 $\\ 
		\hline 
		$J_2$ & $s_{6}^{7}$ & $s_{4}^{7}$ & $s_{6}^{7}$ & $s_{6}^{7}$ & $1.00 $\\ 
		\hline 
		$J_3$ & $s_{6}^{7}$ & $s_{6}^{7}$ & $s_{6}^{7}$ & $s_{6}^{7}$ & $1.00 $\\ 
		\hline 
		$J_4$ & $s_{6}^{7}$ & $s_{6}^{7}$ & $s_{6}^{7}$ & $s_{6}^{7}$ & $1.00 $\\ 
		\hline 
		$J_5$ & $s_{6}^{7}$ & $s_{6}^{7}$ & $s_{6}^{7}$ & $s_{6}^{7}$ & $0.99$\\ 
		\hline 
		$J_6$  & $s_{6}^{7}$ & $s_{6}^{7}$ & $s_{5}^{7}$ & $s_{6}^{7}$ & $1.00 $\\ 
		\hline 
		$J_7$ & $s_{6}^{7}$ & $s_{6}^{7}$ & $s_{6}^{7}$ & $s_{6}^{7}$ & $1.00 $\\ 
		\hline 
		$J_8$  & $s_{6}^{7}$ & $s_{6}^{7}$ & $s_{6}^{7}$ & $s_{6}^{7}$ & $1.00  $\\ 
		\hline 
		$J_9$ & $s_{6}^{7}$ & $s_{6}^{7}$ & $s_{6}^{7}$ & $s_{5}^{7}$ & $0.90$\\ 
		\hline \hline 
	\end{tabular} 
	\caption{Gathered opinion regarding item $I_{27}$ considering the four linguistic criteria and the numerical one.}
	\label{tab:dataI27round2}
\end{table}

\begin{table}[ht!]
  \centering
	\begin{tabular}{c|c|c|c|c}
		\hline \hline
		Judge & Clarity & Writing & Presence & A.Scale \\ 
		\hline 
		$J_1$ & $(s_{12}^{13}, 0)$ & $(s_{12}^{13}, 0)$ & $(s_{12}^{13}, 0)$  & $(s_{12}^{13}, 0)$ \\ 
		\hline 
		$J_2$ & $(s_{12}^{13}, 0)$ & $(s_{8}^{13}, 0)$ & $(s_{12}^{13}, 0)$ & $(s_{12}^{13}, 0)$ \\ 
		\hline 
		$J_3$ & $(s_{12}^{13}, 0)$ & $(s_{12}^{13}, 0)$ & $(s_{12}^{13}, 0)$ & $(s_{12}^{13}, 0)$\\ 
		\hline 
		$J_4$ & $(s_{12}^{13}, 0)$ & $(s_{12}^{13}, 0)$ & $(s_{12}^{13}, 0)$ & $(s_{12}^{13}, 0)$\\ 
		\hline 
		$J_5$ & $(s_{12}^{13}, 0)$ & $(s_{12}^{13}, 0)$ & $(s_{12}^{13}, 0)$ & $(s_{12}^{13}, 0)$\\ 
		\hline 
		$J_6$  & $(s_{12}^{13}, 0)$ & $(s_{12}^{13}, 0)$ & $(s_{10}^{13}, 0)$ & $(s_{12}^{13}, 0)$\\ 
		\hline 
		$J_7$ & $(s_{12}^{13}, 0)$ & $(s_{12}^{13}, 0)$ & $(s_{12}^{13}, 0)$ & $(s_{12}^{13}, 0)$\\ 
		\hline 
		$J_8$  & $(s_{12}^{13}, 0)$ & $(s_{12}^{13}, 0)$ & $(s_{12}^{13}, 0)$ & $(s_{12}^{13}, 0)$ \\ 
		\hline 
		$J_9$ & $(s_{12}^{13}, 0)$ & $(s_{12}^{13}, 0)$ & $(s_{12}^{13}, 0)$ & $(s_{10}^{13}, 0)$\\ 
		\hline \hline
	\end{tabular} 
  	\caption{Item $I_{27}$ unified assessments as 2-tuples linguistic values in round two.}
    	\label{tab:unifiedI27round2}
\end{table}

\begin{table}[H]
	\centering
	\resizebox{16cm}{!} {
		\begin{tabular}{c|ccccc|ccccc}
			\hline \hline
			& \multicolumn{5}{c}{$1^{er}$ Round} & \multicolumn{5}{c}{$2^{nd}$ Round} \\ 
			\hline \hline
			Item & $IS$ & $CS$ & $CI$ & $RS$ & $RI$ & $IS$ & $CS$ & $CI$ & $RS$ & $RI$\\ 
			\hline 
			$I_1$ & $(s_{5}^{7}, -0.183)$ & $true$ & $0.589$  & $false$ & $0.50$  & $(s_{6}^{7}, -0.370)$ & $true$ & $0.819$  & $true$ & $1.00$\\ 
			\hline 
			$I_2$ & $(s_{5}^{7}, -0,326)$ & $true$ & $0.544$  & $false$ & $0.50$  & $(s_{5}^{7}, 0.478)$ & $true$ & $0.758$  & $true$ & $1.00$\\ 
			\hline 
			$I_3$ & $(s_{4}^{7}, 0.096)$ & $true$ & $0.499$  & $false$ & $0.25$  & $(s_{5}^{7}, 0.438)$ & $true$ & $0.728$  & $true$ & $1.00$\\ 
			\hline 
			$I_4$ & $(s_{5}^{7}, -0.311)$ & $true$ & $0.574$  & $true$ & $0.75$  & $(s_{6}^{7}, -0.361)$ & $true$ & $0.797$  & $true$ & $1.00$\\ 
			\hline 
			$I_5$ & $(s_{5}^{7}, -0.305)$ & $true$ & $0.563$  & $true$ & $1.00$  & $(s_{6}^{7}, -0.199)$ & $true$ & $0.863$  & $true$ & $1.00$\\ 
			\hline 
			$I_6$ & $(s_{5}^{7}, 0.056)$ & $true$ & $0.653$  & $true$ & $1.00$  & $(s_{6}^{7}, -0.272)$ & $true$ & $0.825$  & $true$ & $1.00$\\ 
			\hline 
			$I_7$ & $(s_{5}^{7}, -0.041)$ & $true$ & $0.637$  & $true$ & $1.00$  & $(s_{6}^{7}, -0.161)$ & $true$ & $0.883$  & $true$ & $1.00$\\ 
			\hline 
			$I_8$ & $(s_{5}^{7}, -0.021)$ & $true$ & $0.585$  & $true$ & $1.00$  & $(s_{6}^{7}, -0.219)$ & $true$ & $0.860$  & $true$ & $1.00$\\ 
			\hline 
			$I_9$ & $(s_{5}^{7}, 0.024)$ & $true$ & $0.538$  & $true$ & $0.75$  & $(s_{6}^{7}, -0.180)$ & $true$ & $0.874$  & $true$ & $1.00$\\ 
			\hline 
			$I_{10}$ & $(s_{4}^{7}, 0.359)$ & $false$ & $0.388$  & $true$ & $0.75$  & $(s_{6}^{7}, -0.324)$ & $true$ & $0.784$  & $true$ & $1.00$\\ 
			\hline 
			$I_{11}$ & $(s_{4}^{7}, 0.447)$ & $false$ & $0.431$  & $true$ & $0.75$  & $(s_{6}^{7}, -0.446)$ & $true$ & $0.773$  & $true$ & $1.00$\\ 
			\hline 
			$I_{12}$ & $(s_{5}^{7}, 0.046)$ & $true$ & $0.580$  & $true$ & $1.00$  & $(s_{6}^{7},-0.180)$ & $true$ & $0.866$  & $true$ & $1.00$\\ 
			\hline 
			$I_{13}$ & $(s_{5}^{7}, 0.203)$ & $true$ & $0.659$  & $true$ & $1.00$  & $(s_{6}^{7}, -0.355)$ & $true$ & $0.797$  & $true$ & $1.00$\\ 
			\hline 
			$I_{14}$ & $(s_{5}^{7}, -0.001)$ & $true$ & $0.569$  & $true$ & $0.75$  & $(s_{6}^{7}, -0.208)$ & $true$ & $0.853$  & $true$ & $1.00$\\ 
			\hline 
			$I_{15}$ & $(s_{5}^{7}, -0.382)$ & $true$ & $0.512$  & $false$ & $0.50$  & $(s_{6}^{7}, -0.385)$ & $true$ & $0.843$  & $true$ & $1.00$\\ 
			\hline 
			$I_{16}$ & $(s_{5}^{7}, -0.384)$ & $false$ & $0.472$  & $true$ & $1.00$  & $(s_{6}^{7}, -0.497)$ & $true$ & $0.788$  & $true$ & $1.00$\\ 
			\hline 
			$I_{17}$ & $(s_{4}^{7}, 0.484)$ & $true$ & $0.575$  & $false$ & $0.25$  & $(s_{5}^{7}, 0.403)$ & $true$ & $0.741$  & $true$ & $1.00$\\ 
			\hline 
			$I_{18}$ & $(s_{5}^{7}, -0.130)$ & $true$ & $0.561$  & $true$ & $1.00$  & $(s_{6}^{7}, -0.082)$ & $true$ & $0.932$  & $true$ & $1.00$\\ 
			\hline 
			$I_{19}$ & $(s_{5}^{7}, -0.399)$ & $true$ & $0.556$  & $false$ & $0.50$  & $(s_{6}^{7}, -0.389)$ & $true$ & $0.784$  & $true$ & $1.00$\\ 
			\hline 
			$I_{20}$ & $(s_{5}^{7}, 0.200)$ & $true$ & $0.649$  & $true$ & $0.75$  & $(s_{6}^{7}, -0.244)$ & $true$ & $0.831$  & $true$ & $1.00$\\ 
			\hline 
			$I_{21}$ & $(s_{5}^{7}, 0.258)$ & $true$ & $0.611$  & $true$ & $1.00$  & $(s_{6}^{7}, -0.132)$ & $true$ & $0.899$  & $true$  & $1.00$\\
			\hline
			$I_{22}$ & $(s_{5}^{7}, 0.294)$ & $true$ & $0.686$  & $true$ & $1.00$  & $(s_{6}^{7}, -0.115)$ & $true$ & $0.916$  & $true$  & $1.00$\\
			\hline
			$I_{23}$ & $(s_{4}^{7}, 0.355)$ & $false$ & $0.401$  & $false$ & $0.00$  & $(s_{5}^{7}, 0.468)$ & $true$ & $0.728$  & $true$  & $1.00$\\
			\hline
			$I_{24}$ & $(s_{5}^{7}, -0.014)$ & $true$ & $0.530$  & $true$ & $1.00$  & $(s_{6}^{7}, -0.328)$ & $true$ & $0.803$  & $true$  & $1.00$\\
			\hline
			$I_{25}$ & $(s_{5}^{7},-0.084)$ & $true$ & $0.548$  & $true$ & $1.00$  & $(s_{6}^{7}, -0.266)$ & $true$ & $0.816$  & $true$  & $1.00$\\
			\hline
			$I_{26}$ & $(s_{5}^{7},-0.069)$ & $true$ & $0.581$  & $true$ & $1.00$  & $(s_{6}^{7}, -0.286)$ & $true$ & $0.828$  & $true$  & $1.00$\\
			\hline
			$I_{27}$ & $(s_{5}^{7},-0.369)$ & $false$ & $0.493$  & $false$ & $0.50$  & $(s_{6}^{7}, -0.107)$ & $true$ & $0.907$  & $true$  & $1.00$\\
			\hline
			$I_{28}$ & $(s_{5}^{7},-0.431)$ & $false$ & $0.488$  & $true$ & $0.75$  & $(s_{6}^{7}, -0.306)$ & $true$ & $0.819$  & $true$  & $1.00$\\
			\hline
			$I_{29}$ & $(s_{4}^{7},0.315)$ & $false$ & $0.460$  & $false$ & $0.00$  & $(s_{6}^{7}, -0.231)$ & $true$ & $0.852$  & $true$  & $1.00$\\
			\hline
			$I_{30}$ & $(s_{5}^{7}, -0.079)$ & $true$ & $0.631$  & $true$ & $1.00$  & $(s_{6}^{7}, -0.269)$ & $true$ & $0.831$  & $true$  & $1.00$\\
			\hline
			$I_{31}$ & $(s_{5}^{7},-0.162)$ & $true$ & $0.581$  & $true$ & $0.75$  & $(s_{6}^{7}, -0.269)$ & $true$ & $0.831$  & $true$  & $1.00$\\
			\hline
			$I_{32}$ & $(s_{5}^{7},-0.240)$ & $true$ & $0.538$  & $true$ & $0.75$  & $(s_{6}^{7}, -0.208)$ & $true$ & $0.861$  & $true$  & $1.00$\\
			\hline
			$I_{33}$ & $(s_{5}^{7},-0.398)$ & $false$ & $0.462$  & $true$ & $1.00$  & $(s_{6}^{7}, -0.209)$ & $true$ & $0.871$  & $true$  & $1.00$\\
			\hline
			$I_{34}$ & $(s_{5}^{7}, -0.011)$ & $true$ & $0.630$  & $true$ & $0.75$  & $(s_{6}^{7}, -0.307)$ & $true$ & $0.824$  & $true$  & $1.00$\\
			\hline
			$I_{35}$ & $(s_{4}^{7}, 0.382)$ & $true$ & $0.534$  & $false$ & $0.50$  & $(s_{6}^{7}, -0.292)$ & $true$ & $0.811$  & $true$  & $1.00$\\
			\hline
			$I_{36}$ & $(s_{5}^{7}, 0.041)$ & $true$ & $0.630$  & $true$ & $1.00$  & $(s_{6}^{7}, -0.210)$ & $true$ & $0.866$  & $true$  & $1.00$\\
			\hline
			$I_{37}$ & $(s_{5}^{7}, -0.190)$ & $true$ & $0.619$  & $true$ & $1.00$  & $(s_{6}^{7}, -0.119)$ & $true$ & $0.916$  & $true$  & $1.00$\\
			\hline
			$I_{38}$ & $(s_{5}^{7}, -0.344)$ & $true$ & $0.536$  & $true$ & $0.75$  & $(s_{6}^{7}, -0.304)$ & $true$ & $0.812$  & $true$  & $1.00$\\
			\hline
			$I_{39}$ & $(s_{4}^{7},  0.348)$ & $true$ & $0.526$  & $false$ & $0.25$  & $(s_{6}^{7}, -0.328)$ & $true$ & $0.792$  & $true$  & $1.00$\\
			\hline
			$I_{40}$ & $(s_{5}^{7}, -0.080)$ & $true$ & $0.577$  & $true$ & $1.00$  & $(s_{6}^{7}, -0.123)$ & $true$ & $0.908$  & $true$ & $1.00$\\
			\hline
			$I_{41}$ & $(s_{4}^{7}, 0.437)$ & $true$ & $0.540$  & $false$ & $0.25$  & $(s_{6}^{7}, -0.296)$ & $true$ & $0.813$  & $true$ & $1.00$\\
			\hline
			$I_{42}$ & $(s_{4}^{7}, 0.207)$ & $false$ & $0.478$  & $false$ & $0.00$  & $(s_{6}^{7}, -0.273)$ & $true$ & $0.834$  & $true$ & $1.00$\\
			\hline
			$I_{43}$ & $(s_{5}^{7}, 0.332)$ & $true$ & $0.649$  & $true$ & $1.00$  & $(s_{6}^{7}, -0.096)$ & $true$ & $0.893$  & $true$ & $1.00$\\
			\hline
			$I_{44}$ & $(s_{5}^{7},-0.136)$ & $true$ & $0.572$  & $true$ & $0.75$  & $(s_{6}^{7}, -0.266)$ & $true$ & $0.821$  & $true$ & $1.00$\\
			\hline
			$I_{45}$ & $(s_{5}^{7},-0.349)$ & $false$ & $0.491$  & $true$ & $0.75$  & $(s_{6}^{7}, -0.258)$ & $true$ & $0.805$  & $true$ & $1.00$\\
			\hline \hline
			\multirow{3}{*}{Q} & \multicolumn{5}{c}{$1^{er}$ Round} & \multicolumn{5}{c}{$2^{nd}$ Round} \\ 
            & $CC$ & $CW$ & CP & $CAS$ & $\asocrev{QS}$ & $CC$ & $CW$ & CP & $CAS$ & $\asocrev{QS}$\\ 
   
			& $(s_{5}, -0.164)$ & $(s_{3}, -0.354)$ & $(s_{5}, -0.103)$ & $(s_{5}, -0.283)$ & $\asocrev{(s_5, -0.226)}$ & $(s_{6}, -0.265)$ & $(s_{6}, 0.343)$ & $(s_{6}, -0.290)$ & $(s_{6}, -0.229)$ &  $\asocrev{(s_6, -0.282)}$ \\ 
			\hline \hline 
		\end{tabular} 
	}
	\caption{Moderator compares first and second rounds. 2-tuples linguistic values are expressed under $S^7$.}
	\label{tab:round1round2}
\end{table}

\section{Conclusions}\label{section6}

\revdos{This paper worries about the content validation of questionnaires for b-Learning experiences. It focuses on m-Learning and Flipped Classroom methodologies, paying special attention to out-of-class sessions. Firstly, we presented a questionnaire that allows students to evaluate both previous methodologies. Secondly, we proposed the 2-Tuple Fuzzy Linguistic Delphi method to validate the questionnaire taking into account a very diverse panel of experts, as not all judges are experts with the same depth in all areas. Thirdly, we built a DSS web tool to apply the proposed method in a comfortable way. Finally, we conducted a real experience in Higher Education that manifest the suitability of the proposals.}

\revdos{The main findings of this paper are:}
\begin{itemize}
    \item \revdos{The use of the Telegram app and the Moodle platform is suitable to conduct b-Learning experiences under the m-Learning and Flipped Classroom methodologies.}
    \item \revdos{The flexibility of the proposed 2-Tuple Fuzzy Linguistic Delphi method as a consensus-driven DM through} the use of multigranular linguistic term sets allows for comprehensible information given and comprehensible information consumption.
    \item Tool functionalities \revdos{of the DSS}, such as the trimming options, are very useful to understand the threshold levels of discrepancies, so it is a very useful tool for the moderator who interprets the collective levels of reliance and consensus favorably. 
    \item The final evaluation of the questionnaire \revdos{in the real experience} was \emph{Excellent}, so other teachers can use this questionnaire \revdos{to test b-learning} experiences that combine Flipped Classroom and m-Learning.
\end{itemize}

For future work, \revdos{we will} extend \revdos{the} \emph{2tuple-fuzzy-delphi} DSS to integrate more functionality. This software may serve as a test for different multigranular term sets approaches, and for the proposal of new models for information fusion. We also consider to add user management to this tool in order to cover other areas of the Delphi method, such as the communication between the moderator and judges (for instance by sending the evaluation to the expert panel via e-mail), and to improve its usability. \revdos{Moreover, we} also plan to extend the questionnaire format so that it allows to incorporate unconstrained natural language evaluations. \revdos{We will consider the integration of items whose answer can be free natural language text and incorporate opinion analysis into the DM process \cite{ZUHEROS202122, 9800192}.}

\section{Acknowledgment}

This work was partly supported by the \czm{grant PID2020-119478GB-I00 funded by MCIN/AEI/10.13039/501100011033 and by “ERDF A way of making Europe”.} 
\czm{C. Zuheros is supported by the grant PRE2018-083884 funded by MCIN/AEI/10.13039/501100011033 and by ``ESF Investing in your future''.} 


\bibliographystyle{elsarticle-num} 
\bibliography{2023-2TLFDelphi}   

\czmrev{
\newpage
\appendix
\section{Items of the final questionnaire}
\label{annexA}
\begin{enumerate}
\item The activities posed by the teacher through the videos increased my interest in the contents of the course. Scale to be used: Type A
\item Communication with my classmates through collaborative activities increased my interest towards the contents of the course. Scale to be used: Type A
\item The use of videos has facilitated the exchange of information about  the contents of the course. Scale to be used: Type A
\item The collaborative work has facilitated the exchange of information about the contents of the course. Scale to be used: Type A.
\item The use of videos has facilitated the association of ideas related to the contents of the course. Scale to be used: Type A.
\item The collaborative work has facilitated the association of ideas related to the contents of the course.Scale to be used: Type A.
\item The use of videos has facilitated the application of new ideas related to the contents of the course. Scale to be used: Type A
\item The collaborative work has facilitated the application of ideas related to the course contents. Scale to be used: Type A
\item Working collaboratively, I have been able to express my emotions. Scale to be used: Type A
\item Working collaboratively, I have been able to show gratitude to a member of the team. Scale to be used: Type A
\item Working collaboratively, I have been able to express myself freely and without risk. Scale to be used: Type A
\item I felt comfortable interacting with other members of the team. Scale to be used: Type A
\item Working collaboratively, I have felt a sense of unity with the team. Scale to be used: Type A
\item I felt that my point of view was well recognized by other members of the team. Scale to be used.
\item The videos clearly expressed the contents and organization of the course. Scale to be used: Type A
\item Working collaboratively, I have obtained information about the course contents and organization. Scale to be used: Type A
\item The videos encouraged consultation of course content and external sources to generate knowledge among all. Scale to be used: Type A
\item Working collaboratively has promoted and encouraged knowledge building. Scale to be used: Type A
\item Through the videos, I have been given explicit guidance to focus on the course contents. Scale to be used: Type A.
\item Through collaborative work, I have obtained explicit orientations to focus on the contents of the course. Scale to be used: Type A
\item I am satisfied with the activities proposed in the course. Scale to be used: Type B
\item I am satisfied with the information contributed by my peers. Scale to be used: Type B
\item I am satisfied with the answers I received to my concerns, questions and necessities related to the topics covered in the course. Scale to be used: Type B
\item I am satisfied because I was able to express my concerns, questions and necessities concerning the topics covered in the course. Scale to be used: Type B
\item I am satisfied with the agreements adopted in collaborative activities. Scale to be used: Type B
\item I am satisfied with the highlights made in the course activities.Scale to be used: Type B.
\item I am satisfied with the achievement of the objectives of the course. 
Scale to be used: Type B
\item I am satisfied with the conclusions extracted in the collaborative activities. Scale to be used: Type B
\item I am satisfied because I have been able to express emotions, satisfaction, jokes, ironies, etc. Scale to be used: Type B
\item I am satisfied because I have been able to show gratitude to a member of the group. Scale to be used: Type B
\item I am satisfied because I have been able to express affection to members of the team. Scale to be used: Type B
\item I am satisfied because I have been able to express my concerns, questions and necessities on topics outside the course content. Scale to be used: Type B
\item I am satisfied because I have been able to show my personality in the course. Scale to be used: Type B
\item I am satisfied because in the activities terms such as: we, our team, etc. have been used. Scale to be used: Type B
\item I am satisfied because we have supported each other as members of the team
\item I am satisfied with the organization of the course. Scale to be used: Type B
\item I am satisfied with the methodology used in the course. Scale to be used: Type B
\item I am satisfied because together we have encouraged the debate. Scale to be used: Type B
\item I am satisfied because we have reached consensus among all of us. Scale to be used: Type B
\item I am satisfied with the way the content is presented. Scale to be used: Type B
\item I am satisfied because the debates were focused and refocused when necessary. Scale to be used: Type B
\item I am satisfied because the availability of time and location made it easier for me to perform the activity.Scale to be used: Type B
\item I am satisfied with the communication tools used in the course. Scale to be used: Type A
\item I am satisfied with the videos offered in the course. Scale to be used: Type B
\item I am satisfied with the course. 
Scale to be used: Type B
\end{enumerate}
}

\end{document}